\documentclass[options]{JHEP3}
\title{Higher Grading Conformal Affine Toda Theory and (Generalized) Sine-Gordon/Massive Thirring Duality}
\author{Harold Blas\\
Instituto de F\'\i sica Te\'orica - IFT/UNESP\\
Rua Pamplona 145\\
01405-900  S\~ao Paulo-SP, BRAZIL\\
E-mail: \email{blas@ift.unesp.br}}
\abstract{Some properties of the higher grading integrable generalizations of the conformal affine Toda systems are studied. The fields associated to the non-zero grade generators are Dirac spinors. The effective action is written in terms of the Wess-Zumino-Novikov-Witten (WZNW) action associated to an affine Lie algebra, and an off-critical theory is obtained as the result of the  spontaneous breakdown of the conformal symmetry. Moreover, the off-critical theory presents a remarkable equivalence between the Noether and topological currents of the model. Related to the off-critical model we define a real and local Lagrangian provided some reality conditions are imposed on the fields of the model. This real action model is expected to describe the soliton sector of the original model, and turns out to be the master action from which we uncover the weak-strong phases described by (generalized) massive Thirring and sine-Gordon type models, respectively. The case of any (untwisted) affine Lie algebra furnished with the principal gradation is studied in some detail. The example of $\hat{sl}(n)$\,(n=2,3) is presented explicitly.}
\preprint{...}

\def\lab#1{\label{eq:#1}}
\def\br{\begin{eqnarray}}
\def\er{\end{eqnarray}}
\def\be{\begin{equation}}
\def\ee{\end{equation}}
\def\nn{\nonumber}
\def\lb{\lbrack}
\def\rb{\rbrack}
\def\({\left(}
\def\){\right)}

\relax

\def\PRL#1#2#3{{\sl Phys. Rev. Lett.} {\bf#1} (#2) #3}
\def\NPB#1#2#3{{\sl Nucl. Phys.} {\bf B#1} (#2) #3}

\def\PRD#1#2#3{{\sl Phys. Rev.} {\bf D#1} (#2) #3}

\def\PLB#1#2#3{{\sl Phys. Lett.} {\bf #1B} (#2) #3}
\def\JMP#1#2#3{{\sl J. Math. Phys.} {\bf #1} (#2) #3}

\def\PTP#1#2#3{{\sl Prog. Theor. Phys.} {\bf #1} (#2) #3}

\def\AoP#1#2#3{{\sl Annals Phys.} {\bf #1} (#2) #3}

\def\PR#1#2#3{{\sl Phys. Reports} {\bf #1} (#2) #3}

\def\IJMPA#1#2#3{{\sl Int. J. Mod. Phys.} {\bf A#1} (#2) #3}

\def\JHEP#1#2#3{{\sl JHEP} {\bf #1} (#2) #3}

\def\a{\alpha}
\def\b{\beta}

\def\d{\delta}
\def\D{\Delta}

\def\g{\gamma}

\def\vp{\varphi}

\def\h{ {1\over 2}  }

\def\/{\frac}

\def\l{\lambda}
\def\L{\Lambda}

\def\nonu{\nonumber}

\def\pa{\partial}

\def\ra{\rightarrow}

\def\vp{\varphi}

\def\cgh{{\hat {\cal G}}}

\def\ns{N_{{\bf s}}}

\def\({\Big(}
\def\){\Big)}
\def\[{\Big[}
\def\]{\Big]}

\def\rlx{\relax\leavevmode}
\def\inbar{\vrule height1.5ex width.4pt depth0pt}
\def\IZ{\rlx\hbox{\sf Z\kern-.4em Z}}
\def\IR{\rlx\hbox{\rm I\kern-.18em R}}
\def\IC{\rlx\hbox{\,$\inbar\kern-.3em{\rm C}$}}
\def\IN{\rlx\hbox{\rm I\kern-.18em N}}
\def\IO{\rlx\hbox{\,$\inbar\kern-.3em{\rm O}$}}
\def\IP{\rlx\hbox{\rm I\kern-.18em P}}
\def\IQ{\rlx\hbox{\,$\inbar\kern-.3em{\rm Q}$}}
\def\IF{\rlx\hbox{\rm I\kern-.18em F}}
\def\IG{\rlx\hbox{\,$\inbar\kern-.3em{\rm G}$}}
\def\IH{\rlx\hbox{\rm I\kern-.18em H}}
\def\II{\rlx\hbox{\rm I\kern-.18em I}}
\def\IK{\rlx\hbox{\rm I\kern-.18em K}}
\def\IL{\rlx\hbox{\rm I\kern-.18em L}}
\def\one{\hbox{{1}\kern-.25em\hbox{l}}}
\def\0#1{\relax\ifmmode\mathaccent"7017{#1}%
B        \else\accent23#1\relax\fi}

\begin{document}

%\noindent PACS numbers: 04.20.Cv, 03.65.Ta, 04.80.Cc

\section{Introduction}

The so-called conformal affine Toda models coupled to matter fields (CATM) \cite{matter} constitute the higher grading extensions of the conformal affine Toda field theories (CATFT) and have been recently the subject of detailed investigations . In this context, after spontaneously broken the conformal symmetry by setting a field to a constant it has been defined some effective theories the off-critical affine Toda models coupled to matter fields (ATM). This resembles the known relationship between the CATFT massless theories and the massive affine Toda models (ATFT) by which many properties of the affine Toda models can be more easily tackled and understood by regarding them as the conformal affine Toda models with the conformal symmetry spontaneously broken \cite{laf}. Amongst the integrable theories in $1+1$ dimensions there are the conventional conformal Toda models associated to the finite simple Lie algebras. Closely related to them are the massive Toda models, based on loop algebras, which are obtained by perturbing the conformal Toda models with a term associated to the negative of the highest root such that the integrability is preserved \cite{eguchi}. These models exhibit soliton solutions. This is a different procedure in obtaining a massive theory as breaking the conformal symmetry by setting a field to a constant and furthermore the theories obtained this way can have different properties.         
 
The $\hat{sl}(n)$ ATM theories constitute excellent laboratories to test ideas about confinement and the role of solitons in quantum field theories \cite{nucl, nucl1, tension}, duality transformations interchanging solitons and particles \cite{nucl, jmp, annals}, as well as the reduction processes of the (two-loop) Wess-Zumino-Novikov-Witten (WZNW) theory from which the CATM models are derivable \cite{matter}. It has recently been shown that the $sl(2)$ ATM model describes a confinement mechanism and the low-energy spectrum of QCD$_{2}$ (one flavor and $N$ colors) \cite{tension}.  Recently, some $U(1)$ and $U(1)$ x $U(1)$ charged topological soliton solutions of certain class of affine Toda models have been interpreted as charged dyonic domain walls of 4D $SU(N)$ gauge theory and $N=1$ $SU(N)$ SUSY gauge theory in the large $N$ limit, respectively \cite{sotkov1, cabrera}.

Here we construct many field generalizations of sine-Gordon (GSG) and massive Thirring (GMT) models associated to any (untwisted) affine Lie algebra and based on soliton/particle duality and unitarity. These models are known to describe  dual phases of the same theory, namely a sub-model of the off-critical ATM theory defined by a local and real Lagrangian which is expected to describe the dynamics of the soliton type solutions. Beyond the well known $\hat{sl}(2)$ case the related off-critical $\hat{sl}(n)$ ATM model does not possess a local and real Lagrangian, therefore we resort to a sub-model with well behaved classical solutions and real Lagrangian in which a Noether and topological currents equivalence is inherited from the original ATM model. In \cite{bueno} the authors studied the $\hat{sl}(3)$ CATM soliton solutions and some of their properties up to general $2$-soliton. The local and real Lagrangian describing the soliton sector of the $\hat{sl}(3)$ CATM model was shown to be a master Lagrangian from which the generalized sine-Gordon (GSG) or the massive Thirring (GMT) models are derivable \cite{jmp}. Here we generalize this result at the classical level and provide a full Lie algebraic construction for any (untwisted) affine Lie algebra. In the $\hat{sl}(n)$ L-ATM case it will also be clear the duality exchange of the coupling regimes and the generalized soliton/particle correspondences, which we uncover by providing explicit relationships between the GSG and GMT fields. In this way we give a precise field content of both sectors, namely the correct GMT/GSG duality, first undertaken in \cite{halpern} for $su(n)$ at the quantum level.   

To this end we resort to the master Lagrangian approach \cite{deser} and the Faddeev-Jackiw (FJ) symplectic method \cite{ja} to uncover the GSG and GMT dual sectors, respectively. In the previous works on $\hat{sl}(n)(n=2,3)$ ATM theory \cite{jmp, annals} an extension of the FJ method applied to unravel the GSG sector produced the loss of covariance which was restored by means of a complicated canonical transformation, so, here we overcome this problem applying the master Lagrangian method which also provides a final Lagrangian manifestly invariant under the relevant group symmetry of the scalar sector. In dealing with the GMT sector the FJ method turns out to be more appropriate, in  particular the Darboux's transformation, which is inherent to this approach, simultaneously diagonalize the spinor canonical one-forms, decouple these fields from the Toda fields, and in addition maintains the algebraic structures of the spinor sector for any affine Lie algebra.  

The paper is organized as follows. In section 2 the model is presented, as well as its various properties such as the local symmetries and the remarkable Noether and topological currents equivalence for some special gradation structures. In section 3 we address the problem of the Lagrangian formulation of the CATM system in some suitable parameterization. There exists a local action if some extra fields are introduced in the model and that these fields are related to the original CATM fields through non-local transformations. An off-critical ATM theory is defined by spontaneously broken the conformal symmetry by setting the field $\eta$ to a constant. In section 4 by using the gradation structure properties and the currents equivalence condition as guiding principles we define a local and real Lagrangian sub-model which we call L-ATM. In section 5 we consider the L-ATM model as a master Lagrangian. In section 5.1 the GSG model is obtained by replacing into the master action suitable solutions for the ``matter'' fields and gauge fixing some residual gauge symmetries. In section 5.2 the FJ reduction method is applied to the constrained L-ATM theory in order to obtain the GMT model. In section 6 we verify the GSG/GMT relationship by providing some mappings between the fields of the models in such a way that the L-ATM equations of motion decouple into the relevant equations of motion of the dual models. In section 7 we provide an explicit affine Lie algebraic construction for the GSG/GMT correspondence based on the principal gradation of any  (untwisted) affine Lie algebra. In section 8 the examples of $\hat{sl}(2)$ and $\hat{sl}(3)$ are worked out. The last section is devoted to the conclusions and discussions. The Appendix provides affine Lie algebra useful properties and notations.

\section{The model}

In order to make this paper self-contained, in this section we present a brief review of the construction of the CATM models \cite{matter} making some remarks on the results relevant to our discussions. The so-called two-loop WZNW theory is the generalization of the ordinary WZNW model to the affine case. Its action is the same as that for the usual WZNW model. Therefore, the equations of motion for the two-loop WZNW model are \cite{aratyn}
\br
\label{LR}
\pa_{+}(\pa_{-}\hat{g}\hat{g}^{-1})=0, \,\,\,\,\,\pa_{-}(\hat{g}^{-1}\pa_{+}\hat{g})=0.
\er
where $\pa_{\pm}$ are derivatives w.r.t. the light-cone variables $x_{\pm}= t\pm x$, and $\hat{g}$ is an element of the group $\hat{G}$ formed by exponentiating an untwisted affine Lie algebra $\hat{{\cal G}}$. Consider those mappings which can be represented in a ``Gauss decomposition'' form
\br
\label{gauss}
\hat{g}=N B M,
\er
where N, B and M are generated by the sub-algebras $\hat{\cal{G}}_{+}$, $\hat{\cal{G}}_{0}$ and $\hat{\cal{G}}_{-}$, respectively. Introducing the mappings $K_{L/R}$ as $\pa_{-}\hat{g}\hat{g}^{-1}=NK_{L}N^{-1}$ and  $\hat{g}^{-1}\pa_{+}\hat{g}=M^{-1}K_{R}M$, and taking into account (\ref{LR}) one gets
\br
\label{KLR}
\pa_{-}K_{R}=-\left[K_{R},\, \pa_{-}MM^{-1}\right],\,\,\, \pa_{+}K_{L}=\left[K_{L},\, N^{-1}\pa_{+}N\right].
\er  

The construction follows by reducing the two-loop WZNW model imposing the constraints 
\begin{eqnarray}
\label{constr1}
&(\pa_{-}MM^{-1})_{-l}=B^{-1}E_{-l}B,\,\,(\pa_{-}MM^{-1})_{<-l}=0,\\
\label{constr2}
&(N^{-1}\pa_{+}N)_{l}=BE_{l}B^{-1}, \,\,(N^{-1}\pa_{+}N)_{>l}=0.
\end{eqnarray}

Therefore, $\pa_{-}MM^{-1}\, \in \, \bigoplus \hat{{\cal G}}_{-n}$ and $N^{-1}\pa_{+}N \, \in\, \bigoplus \hat{{\cal G}}_{n}$,\, $1\leq n\leq l-1$.

Next, define 
\br
 F^{-}&\equiv& E_{-l} + \sum_{m=1}^{l-1} F_{m}^{-} = B\pa_{-}MM^{-1}B^{-1}\\
 F^{+}&\equiv& E_{l} + \sum_{m=1}^{l-1} F_{m}^{+} = B^{-1}N^{-1}\pa_{+}N B.
\er

Considering (\ref{gauss}) and taking into account the constraints (\ref{constr1})-(\ref{constr2}), the Eqs. (\ref{KLR}) give rise to the system of equations for $F^{\pm}_{m}$ and $B$
\br
\label{eqn1}
\pa_{+}(\pa_{-} B B^{-1})&=& \[E_{-l}\,,\, BE_{l}B^{-1}\] + \sum_{n=1}^{l-1}\[F_{n}^{-}\,,\, BF_{n}^{+}B^{-1}\]  \\
\label{eqn2}
\pa_{-}F_{m}^{+}  &=&  \[E_{l}\,,\, B^{-1}F_{l-m}^{-}B\] + \sum_{n=1}^{l-m-1}\[F_{n+m}^{+}\,,\, B^{-1}F_{n}^{-}B\]\\
\label{eqn3}
\pa_{+}F_{m}^{-}  &=&-  \[E_{-l}\,,\, BF_{l-m}^{+}B^{-1}\] - \sum_{n=1}^{l-m-1}\[F_{n+m}^{-}\,,\, BF_{n}^{+}B^{-1}\].
\er

Parameterize $B$ as 
\br
\label{para}
B\,=\, b\, e^{\eta Q_{{\bf s}}}\, e^{\nu C},\,\,\,\,\,\,\,\,\,\,b = \mbox{exp}\, \hat{{\cal G}}_{0}^{*},
\er
with\,  $\hat{{\cal G}}_{0}^{*}$\,  being a sub-algebra generated by all elements of $\hat{{\cal G}}_{0}$ other than  $Q_{{\bf s}}$ and $C$ (see Appendix). The fields $\nu$ and $\eta$ lie in the directions which are extensions of the loop algebra and are responsible for making the system conformally invariant (see below). Replace (\ref{para}) into (\ref{eqn1})-(\ref{eqn3}) to get 
\br
\label{eqn11}
\pa_{+}(\pa_{-} b b^{-1}) + \pa_{+}\pa_{-}\nu C \,=\, e^{l \eta} \[E_{-l}\,,\, b E_{l} b^{-1}\] + \sum_{n=1}^{l-1} e^{n \eta} \[F_{n}^{-}\,,\,  b F_{n}^{+} b^{-1}\]  \\
\label{eqn21}
\pa_{-}F_{m}^{+}  \,=\, e^{(l-m)\eta} \[E_{l}\,,\, b^{-1} F_{l-m}^{-} b\] + \sum_{n=1}^{l-m-1}e^{n \eta}  \[F_{n+m}^{+}\,,\, b^{-1}F_{n}^{-}b\]\\
\label{eqn31}
\pa_{+}F_{m}^{-}  \,=\,-  e^{(l-m)\eta} \[E_{-l}\,,\, b F_{l-m}^{+} b^{-1}\] - \sum_{n=1}^{l-m-1}  e^{n \eta}  \[F_{n+m}^{-}\,,\, b F_{n}^{+} b^{-1}\]\\
\label{eqn41}
\pa_{+}\pa_{-} \eta Q_{{\bf s}} \,=\, 0.
\er

The system of equations (\ref{eqn11})-(\ref{eqn41}) defines the so-called {\sl conformal affine Toda model coupled to matter fields} (CATM) \cite{matter}. Let us point out that the data $\{\hat{\cal G},\, Q_{{\bf s}},\, l, \, E_{\pm l}\}$
 completely characterize the model defined above and various interesting physical properties will depend crucially on a given data. 

The system (\ref{eqn11})-(\ref{eqn41}) is invariant under the conformal transformation
\br
\label{conformal}
x_{+} \ra {\hat x}_{+} = f(x_{+}) \, , \qquad 
x_{-} \ra {\hat x}_{-} = g(x_{-}),
\er
with $f$ and $g$ being analytic functions and with the fields transforming 
as
\begin{eqnarray}
\nonumber
\widetilde{b}(\hat{x}_{+}\, , \, \hat{x}_{-}) &=& b(x_{+}\, , \, x_{-});\,\,\,\,\,\,\,
e^{-\widetilde{\nu} (\hat{x}_{+}\, , \, \hat{x}_{-})} \,=\,\( f^{\prime}\)^{\d} \, \( g^{\prime}\)^{\d}
e^{-\nu (x_{+}\, , \, x_{-})} \, ,
\\
\nonumber
e^{-\widetilde{\eta} (\hat{x}_{+}\, , \, \hat{x}_{-})} &=& \( f^{\prime}\)^{\frac{1}{l}} \, \( g^{\prime}\)^{\frac{1}{l}}  e^{-\eta (x_{+}\, 
, \, x_{-})} \, ,\,\,\,\,
\widetilde{F}_{m}^{+} (\hat{x}_{+}\, , \, \hat{x}_{-}) \, = \,  \( f^{\prime}\)^{-1+ \frac{m}{l}} F_{m}^{+} (x_{+}\, , \, x_{-}),
\\
\nonumber
\widetilde{F}_{m}^{-} (\hat{x}_{+}\, , \, \hat{x}_{-}) & =&   \( g^{\prime}\)^{-1+ \frac{m}{l}} F_{m}^{-} (x_{+}\, , \, x_{-}),
\end{eqnarray}
where the conformal weight $\d$, associated to $e^{-\nu}$, is arbitrary. 
The conformal symmetry of the reduced model arises from the WZNW conformal symmetry because the current associated to the grading operator can be used to improve the WZNW energy-momentum tensor allowing the constraints (\ref{constr1})- (\ref{constr2}) to weakly commute with it (see more details in \cite{matter, feher}). 
In this paper we will be mainly interested in those models possessing the following local symmetries 
\br
\nonu
b & \rightarrow& h_{L}(x_{-}) b(x_{+}, x_{-})  h_{R}(x_{+})\\
\label{local}
F_{m}^{+} &\rightarrow& h^{-1}_{R}(x_{+}) F_{m}^{+}(x_{+}, x_{-})  h_{R}(x_{+})\\
\nonu
F_{m}^{-} &\rightarrow& h_{L}(x_{-}) F_{m}^{-}(x_{+}, x_{-})  h^{-1}_{L}(x_{-}).
\er

In fact, the system of Eqs. (\ref{eqn11})-(\ref{eqn41}) is invariant under the above symmetries if the following conditions are supplied 
\br
\label{condi}
h_{R}(x_{+})\, E_{l}\,  h^{-1}_{R}(x_{+})\,=\, E_{l},\,\,\,\,\, h^{-1}_{L}(x_{-})\, E_{-l}\,  h_{L}(x_{-})\,=\, E_{-l},
\er
where $h_{L/R}(x_{\mp})\, \in\, {\cal H}_{0}^{L/R}$, \, ${\cal H}_{0}^{L/R}$ being subgroups of $\hat{G}_{0}$.

One of the remarkable properties of some special class of CATM models away from criticality is the Noether and topological currents equivalence. So, we shall assume $l=N_{\bf s}$ in the next derivation. This is precisely the condition which characterizes the models possessing a $U(1)$ Noether current, which, under certain circumstances, is proportional to a topological current. Next, we briefly describe the construction of the chiral currents associated to the elements $z^{-1}E_{N_{\bf s}}$ and  $z E_{-N_{\bf s}}$ and the Hamiltonian reduction CATM $\rightarrow$ ATM (off-critical) \cite{matter}. The procedure follows by constructing two equations using the Eqs. (\ref{eqn2})-(\ref{eqn3}). To get the first, multiply Eq. (\ref{eqn2}) by $z^{-1}F^{+}_{N_{{\bf s}}-m}$, sum over $m$ and take the trace, and in the same Eq. change $m$ by $N_{{\bf s}}-m$ and then multiply by $z^{-1}F^{+}_{m}$, sum over $m$ and take the trace. Add both of them and this defines the first equation. The second equation is constructed similarly, multiply Eq. (\ref{eqn3}) by $zF^{-}_{N_{{\bf s}}-m}$, sum over $m$ and take the trace, next, take the same Eq. and change $m$ by $N_{{\bf s}}-m$, multiply by $zF^{-}_{m}$, sum over $m$ and take the trace, and add both of them to define the second equation. The two equations can be summarized as
\br
\nonu
\sum_{m=1}^{N_{{\bf s}}-1} \pa_{\mp} Tr\( z^{\mp 1}F^{\pm}_{m}F^{\pm}_{N_{{\bf s}}-m}\) &=& \mp 2 \sum_{m=1}^{N_{{\bf s}}-1} Tr\(z^{\mp 1} E_{\pm N_{{\bf s}}} [F^{\pm}_{m}\,, B^{\mp 1} F^{\mp}_{m} B^{\pm 1}\,]\) + \\
&&
X^{\pm},\label{twoeqns}
\er
where
\br
\nonu
 X^{\pm} &=& \sum_{m=1}^{N_{{\bf s}}-2} \sum^{N_{{\bf s}}-m-1}_{n=1} Tr \(z^{\mp 1}F^{\pm}_{N_{{\bf s}}-m}[F^{\pm}_{n+m}\,,\,  B^{\mp 1} F^{\mp}_{n} B^{\pm 1}]\)\\
&& + \sum_{m=2}^{N_{{\bf s}}-1}  \sum^{m-1}_{n=1} Tr\(z^{\mp 1} F^{\pm}_{m} [F^{\pm}_{N_{{\bf s}}+n-m}\,, B^{\mp 1} F^{\mp}_{n} B^{\pm 1} ] \).
\label{xpm}
\er

Notice that each double sum in $ X^{\pm}$  vanishes separately
\br
X^{\pm}\,=\,0.        
\er 

Next, multiply (\ref{eqn1}) by $z^{-1} E_{N_{{\bf s}}}$ ($z E_{-N_{{\bf s}}}$) and take the trace and compare with (\ref{twoeqns}) to get 
\br
\pa_{\mp}{\cal J}_{\pm}\,=\, 0,
\er 
where 
\br
\label{calcur}
{\cal J}_{\pm}\,=\, \pm Tr \(z^{\mp 1} E_{\pm N_{{\bf s}}} B^{\mp 1}\pa_{\pm} B^{\pm 1} \)- \frac{1}{2} \sum_{m=1}^{N_{{\bf s}}-1} Tr \(z^{\mp 1} F_{m}^{\pm}  F_{N_{{\bf s}}-m}^{\pm}  \).
\er

The conditions 
\br
\label{e0}
z E_{-N_{{\bf s}}}\,=\, \zeta\,  z^{-1} E_{N_{{\bf s}}}\,\equiv \, E_{0} \, \in \mbox{ center of}\,\,\,\,  \hat{{\cal G}}_{0},\,\,\, \zeta = \mbox{const.},
\er
allow us to define the conserved currents
\br
\label{J1}
J_{\pm} &=& \mp \frac{1}{2} \zeta^{\pm} \sum_{m=1}^{N_{{\bf s}}-1} Tr \(z^{\mp 1} F_{m}^{\pm}  F_{N_{{\bf s}}-m}^{\pm}\),\,\,\,\,\,\,\,\, \pa_{\mu} J^{\mu}\,=\, 0,\\
j_{\pm}&=& - Tr  \( E_{0} B^{\mp 1}\pa_{\pm} B^{\pm 1} \),\,\,\,\,\,\,\,\,\,\,\pa_{\mu} j^{\mu}\,=\, 0, \,\,\,\,\,\,\,\,\zeta^{+}\equiv \zeta,\,\,\,\zeta^{-}\equiv 1,
\label{j1}
\er  
which are the Noether and topological currents, respectively.
Therefore, gauge fixing the conformal symmetry by imposing the constraints \cite{matter}:  ${\cal J}_{\pm}\,=\, 0$,  the remarkable equivalence between the two types of currents emerges
\br
\label{equivalence}
{\cal J}_{\pm}\,=\, 0\,\, \Rightarrow\,\,  J_{\pm}\, \sim\, j_{\pm}.
\er

Notice that the above reduction amounts to setting $\eta=$constant. Moreover, it has been shown that the soliton type solutions are in the orbit of the vacuum solution  $\eta=0$ (see \cite{matter, nucl1, annals}).

\section{Effective action}

It is well known that the non-Abelian conformal affine Toda  (CAT) model  and some integrable deformations of $WZNW$ theories are equivalent to specific gauged (two-loop) WZNW models (see e.g. \cite{laf, sotkov, feher}). The higher grading generalizations of the Toda systems associated to finite Lie algebras and their effective actions written in terms of WZNW-type actions were considered in \cite{saveliev}. Our aim is to generalize these effective actions to the case of affine Lie algebras. We write a local action by introducing some extra fields $w^{\pm}$ in the model such that these fields are related to the $B$ and $F^{\pm}$ CATM fields through some non-local transformations. The effective action corresponding to the system of equations (\ref{eqn1})-(\ref{eqn3}) is written as
\br
\nonu
\frac{1}{k} I_{eff}^{(l)}&=& I_{WZW}(B) + \int_{M}\{  <E_{-l}, B E_{l}B^{-1}> + \sum_{m=1}^{l-1} <F_{m}^{-}\,,\, B F^{+}_{m} B^{-1}>\\
&&\nonu
 + <E_{-l}\,,\,e^{-w^{+}} \pa_{+} e^{w^{+}}> + <\sum_{m=1}^{l-1} F_{m}^{-}\,,\, e^{-w^{+}} \pa_{+} e^{w^{+}}> +\\
&&
 <\pa_{-} e^{w^{-}} e^{-w^{-}}\,,\, E_{l}>+  < \pa_{-} e^{w^{-}} e^{-w^{-}}\,,\, \sum_{m=1}^{l-1} F_{m}^{+}> \}
\label{effaction}
\er 
where 
\br
\nonu
I_{WZW}(B)\,=\, \frac{1}{8} \int_{M} d^2x Tr(\pa_{\mu}B \pa^{\mu}B^{-1}) + \frac{1}{12} \int_{D} d^3 x\, \epsilon^{ijk} Tr(B^{-1} \pa_{i}B B^{-1} \pa_{j}B B^{-1} \pa_{k}B).
\er
and $w^{-} \in \, \bigoplus \hat{{\cal G}}_{-n}$,\,\, $w^{+} \in\, \bigoplus \hat{{\cal G}}_{n}$,\, $1\leq n\leq l-1$. In the simplest $l=1$ case the action (\ref{effaction}) reduces to the usual conformal affine Toda system. The $l=2$ case coincides with those obtained in \cite{feher, gervais} for half-integral gradation of a finite Lie algebra. In \cite{saveliev} an analogous expression to (\ref{effaction}) has been presented which includes only the $w^{\pm}$ fields. In the case under consideration here we have included the $F^{\pm}$ fields and regarded the $w^{\pm}$ ones as auxiliary fields.        
 
The Euler-Lagrange equation obtained from (\ref{effaction}) by varying $B$ is precisely the Eq. (\ref{eqn1}). The equations derived from (\ref{effaction}) by varying the fields $F^{+}_{m}$ and $F^{-}_{m}$ become 
\br
\label{auxiliar}
B^{-1} F^{-}_{m} B \,=\, -  \( \pa_{-} e^{w^{-}} e^{-w^{-}}\)_{-m},\,\,\,\,\,\,B F^{+}_{m} B^{-1} \,=\, -  \(e^{-w^{+}} \pa_{+} e^{w^{+}}\)_{m}\,,
\er 
and the equations derived by varying $e^{w^{\mp}}$ are
\br
\label{ww}
\pa_{\mp} F^{\pm}_{m}\,=\,-[E_{\pm l}\,,\,\(\pa_{\mp} e^{\pm w^{\mp}} e^{\mp w^{\mp}} \)_{\mp l \pm m}] - \sum_{n=1}^{l-m-1} [F_{m+n}^{\pm}\,,\,\(\pa_{\mp} e^{\pm w^{\mp}} e^{\mp w^{\mp}} \)_{-n}].
\er 

Substituting the expressions for $w^{\mp}$ from (\ref{auxiliar}) into (\ref{ww}) one gets the equations of motion (\ref{eqn2})-(\ref{eqn3}).

Using the parameterization (\ref{para}) and the identities (\ref{identity1})-(\ref{identity2}) one can rewrite the effective action (\ref{effaction}) as
\br
\nonu
\frac{1}{k} I_{eff}^{(l)}&=& I_{WZW}(b) + \int_{M}\{ e^{l \eta} <E_{-l}, b E_{l}b^{-1}> + \sum_{m=1}^{l-1} e^{m \eta} <F_{m}^{-}\,,\, b F^{+}_{m} b^{-1}>\\
&&\nonu
 + <E_{-l}\,,\,e^{-w^{+}} \pa_{+} e^{w^{+}}> + <\sum_{m=1}^{l-1} F_{m}^{-}\,,\, e^{-w^{+}} \pa_{+} e^{w^{+}}> +\\
&&
\nonu
 <\pa_{-} e^{w^{-}} e^{-w^{-}}\,,\, E_{l}>+  < \pa_{-} e^{w^{-}} e^{-w^{-}}\,,\, \sum_{m=1}^{l-1} F_{m}^{+}> \} +\\
&& \frac{1}{4} N_{{\bf s}} \int d^2x \pa_{\mu} \nu \pa^{\mu} \eta.
\label{effaction1}
\er 

Then, the effective action of the CATM system (\ref{eqn11})-(\ref{eqn41}) in the parameterization above becomes the WZW action for the field $b$, plus ``kinetic'' type terms for the fields $ F^{\pm}$ and $w_{m}^{\pm}$, some interaction terms between $F^{\pm}$ and $w^{\pm}$ fields plus a potential involving the terms of type $Tr\(E_{-l}b E_{l}b^{-1}\)$ and $Tr (F_{m}^{+} b F_{m}^{-} b^{-1})$ coupled to $\eta$, plus, finally, a kinetic term for the two fields $\eta$ and $\nu$. The action  (\ref{effaction1}), supplied with  $\eta=\eta_{0}=$const. and the condition (\ref{e0}), defines the effective action of the off-critical {\sl affine Toda model coupled to matter} (ATM) \cite{matter, nucl, annals} and it will describe the Toda field $b$ and the ``matter''  fields $F^{\pm}_{m},\, w^{\pm}_{m}$.

\section{Local affine Toda model coupled to matter (L-ATM)}
\label{ssec:local}

In the following we address the question of locality and reality of the off-critical ATM effective action for the special models in which (\ref{equivalence})  holds; i.e. we consider the action (\ref{effaction1}) with $\eta=$const. and the condition (\ref{e0}).
The relevant physical properties of the ATM model, such as the Noether and topological currents equivalence (\ref{equivalence}) which in turn implies the particle/soliton correspondences, are stated in terms of local expressions of the fields of the model ($b$ and $F^{\pm}_{m}$ ). On the other hand, in some special models associated to the principal gradation of the algebra $\hat{sl}(n)$ and  satisfying (\ref{equivalence}) it has recently been shown, by direct verification of the soliton-type solutions, that the terms in the right hand side of the Eqs. (\ref{eqn21})-(\ref{eqn31}) inside the sums identically vanish for these kind of solutions \cite{jmp, bueno}. Moreover, the theory defined without these terms was shown to display the interesting properties of the original off-critical $\hat{sl}(n)$ ATM model; i.e. Noether/topological currents and soliton/particle correspondences \cite{jmp}.  
 
In order to define a relativistic quantum field theory with real and local Lagrangian we follow the ideas developed in the previous works \cite{nucl1, jmp} in which some reality conditions, consistent with the equations of motion of the model, are imposed on the fields. To proceed, let us notice that any integral gradation of an affine Lie algebra \cite{kac} possesses a period (e.g. for the principal gradation, {\sl period}$ \,=\,N_{{\bf s}}\,=\,h\,=\,$ Coxeter number) such that the eigensubspaces, (e.g. the eigensubspaces of grades  $m$ and $N_{{\bf s}}-m$) have the {\sl same structure}. Moreover, subspaces of grade $m$ and $-m$ are always {\sl isomorphic}. By choosing $E_{l}$ and $E_{-l}$ as generators related by such an isomorphism, and in addition considering $l$ to be equal to the period associated to the gradation, i.e. $l \,=\, N_{{\bf s}}$, one obtains the models satisfying the remarkable equivalence (\ref{equivalence}).  
    
According to the discussions above the subspaces $F^{\pm}_{m}$ and $F^{\pm}_{l-m}$ have the same structure, and in addition one can choose a compact real form of the Lie algebra ${\cal G}_{0}^{*}$ (see appendix of Ref. \cite{laf}) then one is able to impose the reality conditions 
\br
\label{isomor}
\widetilde{(F^{\pm}_{m})} = \epsilon\, F^{\pm}_{l-m},\,\,\,\,\,\,\,b^{\dagger}\,=\, b^{-1},\,\,\,\,\,\,\,\,\, \epsilon\,=\,\pm 1,
\er
where `$\,\,\widetilde{.}\,\,$' maps generators with grade, say $(m)$,  to another one with grade $(l-m)$, and in addition it maps the corresponding `matter' fields which lie in the relevant directions to their complex conjugates, and $\dagger$ is the usual hermitian conjugation operation.
We assume that the representation spaces are supplied with scalar product with respect to which 
\br
\label{hermitian}
(H^{m}_{a})^{\dagger}= H^{-m}_{a},\,\, (E_{\a}^{m})^{\dagger} = E_{-\a}^{-m},\,\,\, C^{\dagger}=C,\,\,  D^{\dagger}=D.  
\er 

The equations of motion for the off-critical ATM model satisfy these reality conditions provided that we impose the constraints 
\br
\label{sumconst}
\sum_{n=1}^{l-m-1} e^{n \eta_{0}} \[F_{n+m}^{+}\,,\, b^{-1}F_{n}^{-}b\]\,=\,0;\,\,\,\sum_{n=1}^{l-m-1}  e^{n \eta_{0}}  \[F_{n+m}^{-}\,,\, b F_{n}^{+} b^{-1}\]\,=\,0,
\er
in the right hand side of the Eqs. of motion (\ref{eqn21})-(\ref{eqn31}).

Notice that the expressions (\ref{sumconst}) have been originated from the terms containing
\br
<\pa_{-} {e^{w^{-}}} e^{-w^{-}}\,,\,F^{+}_{m}>\,\,\,\,\mbox{and}\,\,\,\, <F^{-}_{m}\,,\, e^{-w^{+}} \pa_{+} e^{w^{+}}>
\er
in the effective action (\ref{effaction1}), respectively.
 
The action (\ref{effaction1}) could become highly non-local when written only in terms of the fields ($B$ and  $F^{\pm}_{m}$) of the ATM model. A convenient parameterization allows us to describe the real and local part of the action. Let us introduce the $W_{m}^{\pm}$ fields in the real and local part of the off-critical ATM action (\ref{effaction1}) with $\eta=\eta_{0}=$const., such that
\br
\nonu
\frac{1}{k} I_{\mbox{eff}}^{(l)} &=& I_{\mbox{WZW}}(b) + \int_{M} \{ e^{l \eta_{0}} <E_{-l}, b E_{l}b^{-1}> + \sum_{m=1}^{l-1} \Big[  e^{m \eta_{0}} <F_{m}^{-}\,,\, b F^{+}_{m} b^{-1}>\\
&&
\nonu
-\frac{1}{2} <E_{-l}\, , \, [W^{+}_{m}\, , \, \pa_{+} W^{+}_{l-m}]> + <F_{m}^{-}\, , \, \pa_{+} W^{+}_{m}> \\
&&
+\frac{1}{2} <[W^{-}_{m}\,,\, \pa_{-} W^{-}_{l-m}]\,
 , \, E_{l}> + <\pa_{-} W^{-}_{m}
 \, , \, F_{m}^{+} > \Big]\}.
\label{latm}
\er

This defines the reduced real {\sl (local) affine Toda model coupled to matter fields} (L-ATM). Notice that (\ref{latm}) is obtained from (\ref{effaction}) by neglecting the multiple commutators in the expansions of type $\pa_{-} e^{w^{-}}e^{-w^{-}}= \pa_{-} w^{-} + \frac{1}{2} [w^{-},\pa_{-}w^{-}]+...$, such that only the first term is kept in $<\pa_{-} e^{w^{-}} e^{-w^{-}}\,,\, F_{m}^{+}>$ and due to the gradation structure only the second term is maintained in  $ <\pa_{-} e^{w^{-}} e^{-w^{-}}\,,\, E_{l}>$. The same steps are followed for the $w^{+}$ terms.
 
From the action (\ref{latm}) we get the following equations of motion:
\br
\label{auxiliar1}
\pa_{+} W_{m}^{+} &=& - e^{m \eta_{0}} b F^{+}_{m} b^{-1},\,\,\,\, \pa_{-} W_{m}^{-}\,=\, -  e^{m \eta_{0}} b^{-1} F^{-}_{m} b,\\ 
\label{eqnw1}
\pa_{+}(\pa_{-} b b^{-1})&=& e^{l \eta_{0}} \[E_{-l}\,,\, b E_{l}b^{-1}\] + \sum_{n=1}^{l-1} e^{m \eta_{0}} \[F_{n}^{-}\,,\, bF_{n}^{+}b^{-1}\]  \\
\label{eqnw2}
\pa_{-}F_{m}^{+}  &=& - \[E_{l}\,,\, \pa_{-} W^{-}_{l-m}\];\,\,\,\,
\pa_{+}F_{m}^{-} \,=\,  \[E_{-l}\,,\, \pa_{+} W^{+}_{l-m}\].
\er

The Eq. (\ref{eqnw1}) describes the off-critical sector of Eq. (\ref{eqn11}). Besides, taking into account Eqs. (\ref{auxiliar1}) and the equations (\ref{eqnw2}) one can reproduce the Eqs. (\ref{eqn21})- (\ref{eqn31}) without the sum in the right hand sides, respectively. Then, (\ref{latm}) will describe the system\br
\label{eq1}
\pa_{+}(\pa_{-} b b^{-1})& =&  e^{l \eta_{0}} \[E_{-l}\,,\, b E_{l}b^{-1}\] + \sum_{n=1}^{l-1} e^{m \eta_{0}} \[F_{n}^{-}\,,\, bF_{n}^{+}b^{-1}\]\\
\label{eq2}
\pa_{-}F_{m}^{+}  &=& e^{(l-m)\eta_{0}} \[E_{l}\,,\, b^{-1} F_{l-m}^{-} b\],\\
\label{eq3}
\pa_{+}F_{m}^{-}  &=&-  e^{(l-m)\eta_{0}} \[E_{-l}\,,\, b F_{l-m}^{+} b^{-1}\],
\er
written in terms of the fields $b$ and $F^{\pm}$. It is an easy task to show that the equations of motion (\ref{auxiliar1})-(\ref{eqnw2}) [ or (\ref{eq1})-(\ref{eq3})], as well as the  constraints (\ref{sumconst}), supplied with convenient transformations for the $ W_{m}^{\pm}$ fields, are invariant under the left-right (\ref{local}) local symmetries; i.e.,  these symmetries are preserved by the successive reduction processes.

The remarkable properties of the ATM theory is expected to translate into the reduced real and local models. The reason is that the terms in the Eqs. of motion (\ref{eqn21})- (\ref{eqn31}) related to the nonlocal part of the action (\ref{effaction1}), which were set to zero in  (\ref{sumconst})  to  define the L-CATM model (\ref{latm}), do not contribute to the chiral currents used to implement the constraints in order to make the reduction CATM $\rightarrow$ ATM which gives rise to the above remarkable currents equivalence (\ref{equivalence}). In fact, the $X^{\pm}$ sums in (\ref{xpm}) do not contribute to the relations (\ref{twoeqns}) which are used to construct the conserved two chiral currents ${\cal J}_{\pm}$ (\ref{calcur}). Notice that the $X^{\pm}$ sums arise due to the sums in the right hand sides of Eqs. (\ref{eqn2})-(\ref{eqn3}) [or (\ref{eqn21})-(\ref{eqn31})]. These terms were set to zero to define the (real) L-ATM. Therefore, the L-ATM will inherit the remarkable properties of the off-critical ATM effective action. Then, one expects that the (real)  L-ATM model will inherit from the ATM model such properties as the Noether/topological currents and soliton/particle correspondences  which are intimately related.

Actually, they have been verified in particular instances: the subset of soliton type solutions of the Eqs. of motion (\ref{eqn11})-(\ref{eqn41}) of $\hat{sl}(n)$ ($n=2,3$) CATM model in the orbit of the conformal group ($\eta =0$) satisfy the reality conditions (\ref{isomor}) for the  fields $b$ and $F^{\pm}_{m}$ and the Eqs. of motion (\ref{eq1})-(\ref{eq3}) of the (real) L-ATM model, as well as the equivalence (\ref{equivalence}) \cite{nucl, nucl1, jmp, bueno}.

\section{The L-ATM model as a master Lagrangian}
\label{sec:master}

We regard the (L-ATM) theory (\ref{latm}) as a master action from which the {\sl (generalized) sine-Gordon (GSG)} and  {\sl (generalized) massive Thirring (GMT)} dual models are derivable. In the next subsections we follow the 
 master action approach (see e.g. \cite{deser}) and the Faddeev-Jackiw (FJ) method \cite{ja} to unravel the GSG and GMT dual sectors, respectively.

\subsection{Generalized sine-Gordon model (GSG)}
\label{ssec:gsg}

In order to uncover the GSG sector of the model (\ref{latm}) we follow the well known technique to perform duality mappings using the master action approach \cite{deser}. Although this technique was initially applied to vectorial models it will prove to be useful in our context. This technique consists of two points. (1) provide a master action describing the model and some auxiliary fields. (2) establish a field equation for the auxiliary fields, solve a subset of fields in terms of the other fields and then substitute the solution into the master action. The actions are equivalent at the classical level and the relationship between them is usually referred as duality. In connection to these points, recently a parent Lagrangian method was used to give a generalization of the dual theories concept for non $p$-form fields \cite{casini}. In \cite{casini}, the parent Lagrangian contained both types of fields, from which each dual theory was obtained by eliminating the other fields through the equations of motion. \footnote{Proving duality by substituting one set of solutions of field equations into the master action would be not legitimate. The correct strategy is to compare the field equations corresponding to the dual theories. The procedure will be justified below (Section \ref{sec:ws}) when we compare the field equations of both dual models through some mapping between the spinors (Dirac fields) and the  scalars (Toda fields).} 

Following the approach one can remove the derivatives over $x_{\pm}$ in Eqs. (\ref{eqnw2}) and write them as 
\br
\label{eqnw21}
F_{m}^{+}  \,=\, - \[E_{l}\,,\, W^{-}_{l-m}\] - f_{m}^{+}(x_{+}),\,\,\,\,
F_{m}^{-}  \,=\,  \[E_{-l}\,,\,  W^{+}_{l-m}\] +  f_{m}^{-}(x_{-}),
\er
with $f_{m}^{\pm}(x_{\pm})$ being analytic functions. Making use of Eqs. (\ref{eqnw21}) and (\ref{auxiliar1}) and disregarding surface terms one can write the master action (\ref{latm}) as
\br
\nonu
\frac{1}{k} I_{eff}^{(l)}&=& I_{WZW}(b) + \int_{M} \{ e^{l \eta_{0}} <E_{-l}b E_{l}b^{-1}> -\frac{1}{2} \sum_{m=1}^{l-1}  e^{m \eta_{0}} \Big[ 2<f_{m}^{-} b f_{m}^{+}  b^{-1}> +\\
&& <[E_{-l}\,,\,W^{+}_{l-m}] b f_{m}^{+}  b^{-1}>+ <[E_{l}\,,\,W^{-}_{l-m}] b^{-1} f_{m}^{-} b> \Big]\}
\label{ffs}
\er

Solving the auxiliary fields $f_{m}^{\pm}$ by means of their Euler-Lagrange equations of motion and replacing into the action (\ref{ffs}) one gets
\br
\nonu
\frac{1}{k} I_{eff}^{(l)}&=& I_{WZW}(b) + \int_{M} \{ e^{l \eta_{0}} <E_{-l}b E_{l}b^{-1}> +\\
&& \frac{1}{4} \sum_{m=1}^{l-1}  e^{m \eta_{0}} < [E_{l}\,,\, W_{l-m}^{-}] b^{-1} [E_{-l}\,,\, W_{l-m}^{+}]  b>\}.
\label{ffs1}
\er

In order to obtain the GSG sector we resort to a convenient gauge fixing procedure. In fact, the action (\ref{ffs1}) is invariant under the local symmetries (\ref{local}) provided that  $W^{-}_{m} \rightarrow h^{-1}_{R}(x_{+}) W_{m}^{-}  h_{R}(x_{+})$,\,\,\, $W_{m}^{+} \rightarrow h_{L}(x_{-}) W_{m}^{+}  h^{-1}_{L}(x_{-})$, and $E_{\pm l}$ satisfy (\ref{condi}). Therefore, one can make the transformation:  $b \rightarrow b'$ and $[E_{\mp l}\,,\,W^{\pm}_{l-m}] \rightarrow \Lambda^{\mp}_{l-m}$, with $\Lambda^{\pm}_{l-m}$ being constant elements in $\hat{{\cal G}}_{\pm (l-m)}$ satisfying reality conditions similar to $F^{\pm}_{l-m}$ in (\ref{isomor}). 
Then the gauge fixed version of (\ref{ffs1}) becomes (set $\eta_{0}=0$) 
\br
\frac{1}{k} I_{eff}^{(l)}&=& I_{WZW}(b) + \int_{M} \[ <E_{-l}b E_{l}b^{-1}> + \sum_{m=1}^{l-1}  <\Lambda_{m}^{-} b \Lambda_{m}^{+} b^{-1}>\].
\label{gsg1}
\er
Below we shall consider the equation of motion 
\br
\label{gsg}
\pa_{+}(\pa_{-} b b^{-1})= \sum_{n=1}^{l-1} \[\L_{n}^{-}\,,\, b \L_{n}^{+}b^{-1}\]
\er
which is obtained from (\ref{gsg1}) in the special case (\ref{e0}); i.e $[E_{\pm l} \, ,\, {\cal G}_{0}^{*}]\,=\,0$. The Eq. (\ref{gsg}) defines the {\sl generalized sine-Gordon model (GSG)}. The parameters $\Lambda^{\pm}_{m}$ have dimension of mass and $k\,=\, \kappa/2\pi$,\, ($\kappa \in \IZ$). The model (\ref{gsg1}) for $l=1$ and particular $E_{l}$'s  has been considered in the literature from different points of view; e.g. in \cite{sotkov} the soliton and solitonic string spectrum, of a version with non-Abelian ${\cal G}_{0}^{*}$ and some local symmetries, have recently been obtained and the relevance of its charge spectrum in the ``off-critical'' $AdS_{3}/CFT_{2}$ correspondence have been discussed. In the next sections we will relate the model (\ref{gsg}) to a generalized massive Thirring model.

\subsection{Generalized massive Thirring model (GMT)}
\label{ssec:gmt}

In this section we perform the reduction using the Faddeev-Jackiw (FJ) method which proved to be useful in the simplest case of $\hat{sl}(n)$ ($n=2,3$) L-ATM models \cite{annals, jmp}. This method will allow us to uncover the GMT sector of the master action (\ref{latm}). To this end we use some constraints of the type (\ref{equivalence}) which are implemented through Lagrange multipliers and  considered as additional terms in the master action. Let us write the relationship (\ref{equivalence}) and the master action in terms of the $b$ and $W^{\pm}_{m}$ fields by using the solution  (\ref{eqnw21}) with $f^{\pm}_{m}=0$. Then from (\ref{eqnw21}) and (\ref{J1})-(\ref{equivalence}) one has
\br
\label{e1}
Tr\[E_{0}\(b^{-1} \pa_{+}b -\frac{1}{2\zeta} \sum_{m}[W_{l-m}^{-}\,,\, [E_{l}\,,\, W_{m}^{-}]]\) \]\,=\,0,
\er
and 
\br
\label{e2}
Tr\[E_{0}\(b \pa_{-}b^{-1} + \frac{1}{2} \sum_{m}[W_{l-m}^{+}\,,\, [E_{-l}\,,\, W_{m}^{+}]]\) \]\,=\,0.
\er

Substituting $F^{\pm}_{m}$ from (\ref{eqnw21}) (take $f^{\pm}_{m}=0$ for simplicity) into (\ref{latm}) we get
\br
\nonu
\frac{1}{k} I_{\mbox{eff}}^{(l)} &=& I_{\mbox{WZW}}(b) + \int_{M} \{ <E_{-l} b E_{l}b^{-1}> -\\
\nonu
&& \sum_{m=1}^{l-1} \Big[ <[E_{-l}\,,\, W_{m}^{+}] b [E_{l}\,,\, W_{m}^{-}] b^{-1}>+\\
\nonu
&&\frac{1}{2} <[E_{-l}\, , \, W^{+}_{m}] \pa_{+} W^{+}_{l-m}> - \frac{1}{2} <[ E_{l}\,,\, W^{-}_{m}] \pa_{-} W^{-}_{l-m}] > \Big] - \\
&&
<\l_{-} \( \pa_{+}b b^{-1} + \frac{1}{2} b \sum_{m} \mu_{m} [[E_{l}\,,\, W_{m}^{-}]\,,\, W_{l-m}^{-}] b^{-1}\)> +\nonu
\\
&&
 <\l_{+} \( \pa_{-}b b^{-1} + \frac{1}{2} \sum_{m} \nu_{m} [[E_{-l}\,,\, W_{m}^{+}]\,,\, W_{l-m}^{+}] \)>\},
\label{latmcons}
\er
where we have included some constraints of the type (\ref{e1})-(\ref{e2}) through Lagrange multipliers $\l_{\pm} \in \hat{{\cal G}}_{0}$, $\l_{\pm}=\l_{\pm}^{a} T_{a}^{(0)}$. Clearly, $\l_{a}$ will stand for $Tr \(T_{a}^{(0)} \l \)$. We have included the parameters $\mu_{m}$ and $\nu_{m}$, they will be fixed below to write a covariant final Lagrangian. Here the sum in ``$m$'' is understood  as follows: for each spinor bilinear associated to the positive root $\a^{(m)}$ with height ``$m$''  and its degeneracy denoted by ``$i$'' one must associate a parameter $\mu_{m}^{i}$ ($\nu_{m}^{i}$) in front of it.  

The L-ATM system (\ref{latm}) possesses global versions of the symmetries (\ref{local}) for subgroups such that $h_{L}=h_{R}^{-1}=h_{D}=$constant, namely 
\br
b \rightarrow h_{D} b   h_{D}^{-1},\,\,\,\, F^{\pm}_{m} \rightarrow  h_{D}  F^{\pm}_{m} h_{D}^{-1} ,\,\,\,\,\,\, W^{\pm}_{m} \rightarrow   h_{D}^{-1} W^{\pm}_{m}  h_{D},
\er  
and taking into account the relationships (\ref{eqnw21}) with $f^{\pm}_{m}=0$, these symmetries give rise to $U(1)$\, Noether currents $(j^{\mu})_{a, m}$ with components
\br
\label{u1}
(j^{0})_{a; m,i}&=&\frac{1}{8} \([[E_{-l}\,,\, W_{m,i}^{+}]\,,\, W_{l-m, i}^{+}] -
 [[E_{l}\,,\, W_{m, i}^{-}]\,,\, W_{l-m, i}^{-}] \)_{a}\\
\label{u11}
(j^{1})_{a; m, i}&=&\frac{1}{8} \([[E_{-l}\,,\, W_{m, i}^{+}]\,,\, W_{l-m, i}^{+}] + [[E_{l}\,,\, W_{m, i}^{-}]\,,\, W_{l-m, i}^{-}]\)_{a},
\er
with possible degenerations for each $m$ denoted by the index $i$. So, we have the conservation laws
\br
\label{currents}
\pa_{\mu} j^{\mu}_{a; m, i} \,=\, 0.
\er

In the case of the principal gradation (see below) $h_{D}=\mbox{exp}({\cal H})$, ${\cal H}=$ Cartan sub-algebra of ${\cal G}_{0}$, there are $N$(=number of positive roots) $U(1)$ currents.

In order to apply the FJ method we need to write (\ref{latmcons}) as a first order Lagrangian. Let us parameterize $b=b(\xi^{a})$, $a = 1,...,dim({{\cal G}_{0}^{*}})$. To find the canonical momenta we introduce the two-form $ {\cal A}\,=\, \frac{1}{2} A_{ab}(\xi)\, d\xi^{a} \wedge d\xi^{b}$ which allows as to rewrite the Wess-Zumino term as
\br  
\frac{1}{3} Tr \( db\, b^{-1}\)^{3}\,=\,d  {\cal A}.
\er

Fortunately we will not need to know ${\cal A}$  explicitly below, and in some special cases this term indeed vanishes, as is the case for Abelian ${{\cal G}_{0}^{*}}$.
We next introduce the nonsingular matrix $ N_{ab}(\xi)$ by
\br
\label{nab}
\frac{\pa b}{\pa \xi^{a}} b^{-1} = N_{ab}(\xi) T^{b}, \,\,\,\,\,\,\,\,T^{b} \in {{\cal G}_{0}^{*}}.
\er

Then the canonical momentum conjugate to  $\xi^{c}$ is written as
\br
\Pi_{c}\, =\,  \frac{\pa {\cal L}}{\pa \dot{\xi}^{c}}\,=\, \frac{1}{4} \( N_{cb} N_{db} \dot{\xi}^{d} - A_{ca} \pa_{x} \xi^{a}\) -\frac{1}{2} \l_{-}^{a}N_{ca} + 
\frac{1}{2} \l_{+}^{a} N_{ca}.
\er

On the other hand, let us write the spinor fields as
\br
\label{param}
W_{m}^{+}\,=\,\sum_{\l(m), i} {\psi}_{L}^{\l(m), i}  T^{(m)}_{\l(m), i},\,\,\,\,
W_{m}^{-}\,=\,\sum_{\l(m), i} {\psi}_{R}^{\l(m), i}  T^{(-l+m)}_{\l(m), i}
\er 
where $ T^{(m)}_{\l(m), i}$ and  $ T^{(-l+m)}_{\l(m), i}$ are the eigenvectors of $[E_{-l}\,,\,[E_{l}\,,\, . ]]$ with nonzero eigenvalues $\l(m)$, and the index $i$ stands for a possible degeneracy of  $\l(m)$.

We are assuming that the Dirac fields are {\sl anti-commuting Grasmannian} variables and their momenta variables defined through {\bf left} derivatives.
The momenta conjugated to the spinors ${\psi}_{R, L}^{\l(m), i}$ are found to be   
\br
\Pi_{m,i}^{+}\, =\, \frac{\pa {\cal L}}{\pa \dot{\psi}_{R}^{\l(m), i}}\,=\, \frac{1}{4} <[E_{-l}\,,\,W^{+}_{l-m}] T^{(m)}_{\l(m), i}>,\\
\Pi_{m,i}^{-}\, =\, \frac{\pa {\cal L}}{\pa \dot{\psi}_{L}^{\l(m), i}}\,=\,  -\frac{1}{4} <[E_{l}\,,\,W^{-}_{m}] T^{(-l+m)}_{\l(m), i}>,
\er
The Hamiltonian of the theory (\ref{latmcons}) is then given by $H=\int dx {\cal H} $ with  
\br
\label{hamilton1}
{\cal H} &=& \Pi_{c} \dot{\xi}^{c} +\Pi_{m,i}^{+} \dot{W}^{+}_{m,i}+ \Pi_{m,i}^{-} \dot{W}^{-}_{m,i}- {\cal L}
\er
or explicitly
\br
\nonu
{\cal H} &=& 2 ( \Pi_{c} + \frac{1}{4} A_{cb} \pa_{x} \xi^{b}) N^{-1}_{ac}  N^{-1}_{ad} ( \Pi_{d} + \frac{1}{4} A_{de} \pa_{x} \xi^{e})+ \frac{1}{8}(\pa_{x}b b^{-1})^{2}\\\nonu
&&
+ 4  N^{-1}_{ab} ( \Pi_{b} + \frac{1}{4} A_{bc} \pa_{x} \xi^{c}) \l_{1}^{a}+ 
2 \l_{1}^{a} \l_{1}^{a}-  <E_{-l} b E_{l}b^{-1}> +\\
\nonu
&&
 \sum_{m=1}^{l-1} \( <[E_{-l}\,,\, W_{m}^{+}] b [E_{l}\,,\, W_{m}^{-}] b^{-1}>-\frac{1}{4} <[E_{-l}\, , \, W^{+}_{l-m}] \pa_{x} W^{+}_{m}> -
\\\nonu
&& \frac{1}{4} <[ E_{l}\,,\, W^{-}_{l-m}] \pa_{x} W^{-}_{m}]> \) + \l_{0}^{a}N_{ba}\pa_{x}\xi^{b} +\\
&&\nonu
 \frac{1}{2} <\l_{0}\, \sum_{m=1}^{l-1}\( b \mu_{m} [[E_{l}\,,\, W_{m}^{-}]\,,\, W_{l-m}^{-}] b^{-1} -\nu_{m} [[E_{-l}\,,\, W_{m}^{+}]\,,\, W_{l-m}^{+}] \)>-\\
&&
\frac{1}{2}  <\l_{1}\,\sum_{m=1}^{l-1} \( b \mu_{m} [[E_{l}\,,\, W_{m}^{-}]\,,\, W_{l-m}^{-}] b^{-1} + \nu_{m} [[E_{-l}\,,\, W_{m}^{+}]\,,\, W_{l-m}^{+}] \)>.\nonu
\\
\label{hamilton}
\er

As usual the first order Lagrangian is obtained from (\ref{hamilton1}) simply by written 
\br
\label{lag1}
{\cal L}&=& \Pi_{c} \dot{\xi}^{c} +\Pi_{m,i}^{+} \dot{W}^{+}_{m,i}+ \Pi_{m,i}^{-} \dot{W}^{-}_{m,i}- {\cal H}
\er
where ${\cal H}$ is given by (\ref{hamilton}).
The next step in the FJ method consists in writing the equations of motion for the Lagrange multipliers. These equations allow us to solve the $\l_{1}^{a}$'s
\br
 \l_{1}^{a}\,=\,- N^{-1}_{ab} ( \Pi_{b} + \frac{1}{4} A_{bc} \pa_{x} \xi^{c})+
\frac{1}{8} \sum_{m=1}^{l-1} \( b \frac{1}{\mu} [[E_{l}\,,\, W_{m}^{-}]\,,\, W_{l-m}^{-}] b^{-1} +\nonu
\\ \frac{1}{\nu} [[E_{-l}\,,\, W_{m}^{+}]\,,\, W_{l-m}^{+}] \)^{a}
\er
and the remaining equations for $\l_{0}^{a}$ lead to the constraints
\br
\label{const1}
2 N_{ba}\pa_{x}\xi^{b}\,=\, \sum_{m}^{l-1} \(\nu_{m} [[E_{-l}\,,\, W_{m}^{+}]\,,\, W_{l-m}^{+}]-\mu_{m} b  [[E_{l}\,,\, W_{m}^{-}]\,,\, W_{l-m}^{-}] b^{-1} \)_{a}
\er

The Lagrange multipliers $\lambda^{a}_{1}$ must be 
replaced back in (\ref{lag1}) and the constraints (\ref{const1}) solved for the $b$ field in terms of the $ W^{\pm}$ fields. Firstly, 
let us replace the $\lambda^{a}_{1}$ multipliers into (\ref{lag1}) to get
\br
\nonu
{\cal L}\,=\,  \sum_{m=1}^{l-1} \( \frac{1}{4} <[E_{-l}\, , \, W^{+}_{l-m}]\dot{W}^{+}_{m}> - \frac{1}{4} <[ E_{l}\,,\, W^{-}_{l-m}] \dot{W}^{-}_{m}]> +\\
\nonu
 \frac{1}{4} <[E_{-l}\, , \, W^{+}_{l-m}]\pa_{x} W^{+}_{m}> + \frac{1}{4} <[ E_{l}\,,\, W^{-}_{l-m}] \pa_{x} W^{-}_{m}]>-
\\
\nonu 
<[E_{-l}\,,\, W_{m}^{+}] b [E_{l}\,,\, W_{m}^{-}] b^{-1}>\) + \Pi_{a} \dot{\xi}^{a}- \Pi_{a} (\hat{J}^{1})^{a} +\\
 \frac{1}{8} \sum_{m,n=1}^{l-1}< \mu_{m} \nu_{n} [[E_{l}\,,\, W_{m}^{-}]\,,\, W_{l-m}^{-}] b^{-1} [[E_{-l}\,,\, W_{n}^{+}]\,,\, W_{l-n}^{+}] b> 
\label{lag2}
\er
where
\br
\nonu
(\hat{J}^{1})^{a} &=& \frac{1}{2} \sum_{m=1}^{l-1} \{ \nu_{m} [[E_{-l}\,,\, W_{m}^{+}]\,,\, W_{l-m}^{+}] +  \mu_{m}
b [[E_{l}\,,\, W_{m}^{-}]\,,\, W_{l-m}^{-}] b^{-1} \}^{a}
\er

Before solving the constraints (\ref{const1}) it is convenient to make the following Darboux's transformations
\br
W_{m}^{-} \Rightarrow b^{-1/2}  W_{m}^{-}  b^{1/2},\,\,\,\, W_{m}^{+} \Rightarrow b^{1/2}  W_{m}^{+}  b^{-1/2}
\er
in (\ref{const1}) and (\ref{lag2}). This procedure is used in the FJ method in order to diagonalize the canonical one-forms. Then (\ref{lag2}) turns out to be
\br
\nonu
{\cal L}\,=\,  \sum_{m=1}^{l-1} \( \frac{1}{4} <[E_{-l}\, , \, b^{1/2} W^{+}_{l-m} b^{-1/2}]\pa_{t}( b^{1/2} W^{+}_{m} b^{-1/2})> - \\
\nonu
\frac{1}{4} <[ E_{l}\,,\, b^{-1/2} W^{-}_{l-m} b^{1/2}] \pa_{t}( b^{-1/2} W^{-}_{m} b^{1/2})]> +\\
\nonu
 \frac{1}{4} <[E_{-l}\, , \, b^{1/2} W^{+}_{l-m}  b^{-1/2}]\pa_{x}(  b^{1/2}W^{+}_{m} b^{-1/2})> +
\\
\nonu
 \frac{1}{4} <[ E_{l}\,,\,  b^{-1/2} W^{-}_{l-m} b^{1/2}] \pa_{x}( b^{-1/2} W^{-}_{m}  b^{1/2})]>-
\\
\nonu 
 <[E_{-l}\,,\, W_{m}^{+}] [E_{l}\,,\, W_{m}^{-}] >\) + \Pi_{a} \dot{\xi}^{a}- \Pi_{a} (J^{1})^{a} +
\\
 \frac{1}{8} \sum_{m,n=1}^{l-1}< \mu_{m} \nu_{n} [[E_{l}\,,\, W_{m}^{-}]\,,\, W_{l-m}^{-}] [[E_{-l}\,,\, W_{n}^{+}]\,,\, W_{l-n}^{+}] > 
\label{lag3}
\er
where $(J^{1})^{a}\,=\,  \frac{1}{2} \sum_{m} \{ \nu_{m} [[E_{-l}\,,\, W_{m}^{+}]\,,\, W_{l-m}^{+}] +  \mu_{m}
 [[E_{l}\,,\, W_{m}^{-}]\,,\, W_{l-m}^{-}] \}^{a}$. The constraints (\ref{const1}) become
\br
\label{const2}
2 N_{ba}\pa_{x}\xi^{b}\,=\, \sum_{m}^{l-1} \(\nu_{m} [[E_{-l}\,,\, W_{m}^{+}]\,,\, W_{l-m}^{+}]- \mu_{m} [[E_{l}\,,\, W_{m}^{-}]\,,\, W_{l-m}^{-}] \)_{a}
\er

From this point forward we will concentrate on models such that
\br
\label{abelian}
[E_{\pm l}\,,\,{\cal G}_{0}^{*}]\,=\,0,\,\,\,\,\,\,[{\cal G}_{0}^{*}\,,\,{\cal G}_{0}^{*}]=0.
\er

The Abelian nature of the sub-algebra ${\cal G}_{0}^{*}$ allows us to write $N_{ab}(\xi)\,=\,1$ in (\ref{nab}). This condition turns out the equations for $\dot{\xi}^{a}$, obtained by taking the time derivative of the constraints (\ref{const2}), more tractable; in fact, these can be solved for $\dot{\xi}^{a}$ by removing one $x$ derivative on both sides of the equations which appear once the conservation laws (\ref{currents}) are taken into account. For non-Abelian ${\cal G}_{0}^{*}$ solving these constraints proves to be a difficult task due to the nontrivial dependence of $N_{ab}(\xi)$ on the fields $\xi$.

The next step consists in writing $\dot{\xi}^{a}$ and $\pa_{x} {\xi}^{a}$ which appear in the kinetic part of (\ref{lag3}) in terms of the $W_{m}^{\pm}$ fields by making use of the constraints (\ref{const2}), the conditions (\ref{abelian}) and the conservation laws (\ref{currents}) which relates $\pa_{t} j^{0}$ to $\pa_{x} j^{1}$. This procedure will give rise to current-current interaction terms in (\ref{lag3}) and taking $\mu_{m}=\nu_{m}$ one obtains a covariant Lagrangian. Notice that the terms containing the $\Pi_{a}$'s in Eq. (\ref{lag3}) cancel to each other if one uses the current conservation laws and the constraints (\ref{const2}). Thus, one is left with the {\sl generalized massive Thirring model (GMT)}  
\br
\nonu
{\cal L}_{GMT}\,=\,  \sum_{m=1}^{l-1} \( \frac{1}{2} <[E_{-l}\, , \, W^{+}_{l-m}]\pa_{+} W^{+}_{m}> - 
\frac{1}{2} <[ E_{l}\,,\,  W^{-}_{l-m}] \pa_{-} W^{-}_{m}]>-\\
 <[E_{-l}\,,\, W_{m}^{+}] [E_{l}\,,\, W_{m}^{-}] >\)-
 \frac{1}{2}  \sum_{m, n} < g_{mn} (j^{+})_{ m}\,  (j^{-})_{ n}>, 
\label{gmt}
\er
where $g_{mn} = 4 \(\mu_{m} \mu_{n}+\mu_{m}+\mu_{n}\)$ and the currents are given in (\ref{u1})-(\ref{u11}). The coupling parameters $g_{mn}$ are symmetric by construction, then the current-current interaction terms can be put in a manifestly covariant form.
 The GMT model (\ref{gmt}) includes the kinetic terms, the mass terms $<[E_{-l}\,,\, W_{m}^{+}] [E_{l}\,,\, W_{m}^{-}] >$, and the general covariant current-current interaction terms.  The canonical pairs are 
\br
\nonu
([E_{-l}\, , \, W^{+}_{l-m}]_{i} \,,\,  W^{+}_{m, i})\,\,\,\, \mbox{and}\,\,\,\, ([ E_{l}\,,\,  W^{-}_{l-m}]_{i} \,,\, W^{-}_{m , i}),
\er 
where $ W^{\pm}_{m , i} \equiv$ Tr $\( W^{\pm}_{m}\, T^{(\mp m)}_{\l (m), i}\)$\,;\, $ [E_{-l}\, , \, W^{+}_{l-m}]_{i} \equiv$ Tr $\( [E_{-l}\, , \, W^{+}_{l-m}]\, T^{(\pm m)}_{\l (m), i}\)$.

From (\ref{gmt}) one gets the GMT Eqs. of motion
\br
\nonu
\pa_{+}F_{m}^{-} \,=\,[E_{-l}\,,\,\pa_{+} W^{+}_{l-m}]&=& [E_{-l}\,,\,[E_{l}\,,W_{m}^{-}]]+\\
&&\frac{1}{16} [[E_{-l}\,,\, W^{+}_{l-m}]\,,\sum_{n=1}^{l-1} g_{mn} [[E_{l}\,,\, W^{-}_{n}]\,,\,  W^{-}_{l-n}\,]]\label{eqt1}
\\
\nonu
-\pa_{-}F_{m}^{+} \,=\,[E_{l}\,,\,\pa_{-} W^{-}_{l-m}]&=&- [E_{l}\,,\,[E_{-l}\,,W_{m}^{+}]]-\\
&&\frac{1}{16} [[E_{l}\,,\, W^{-}_{l-m}]\,,\sum_{n=1}^{l-1} g_{mn}  [[E_{-l}\,,\, W^{+}_{n}]\,,\,  W^{+}_{l-n}\,]]
\label{eqt2}
\er

\section{GSG/GMT relationship}
\label{sec:ws}

Here we provide the GSG/MTM correspondence by decoupling the L-ATM Eqs. of motion (\ref{eq1})-(\ref{eq3}) \( set $\eta_{0}=0$ in Eqs. (\ref{sumconst}) and take $[E_{-l}\,,\, b E_{l}b^{-1}]=0$\, in Eq. (\ref{eq1})\) into the Eqs. of motion of the GSG (\ref{gsg}) and GMT (\ref{eqt1})-(\ref{eqt2}) models, respectively. In fact, consider the relationships
\br
\label{map1}
\sum_{n=1}^{l-1} \[F_{n}^{-}\,,\, bF_{n}^{+}b^{-1}\]  &= & \sum_{n=1}^{l-1}\[\L_{n}^{-}\,,\, b\L_{n}^{+}b^{-1}\] ,\\
\label{map2}
\[E_{-l}\,,\, bF_{l-m}^{+}b^{-1}\] &= &  \[E_{-l}\,,\,F_{l-m}^{+}\]+\frac{1}{16} \[\sum_{n} g_{mn} \[ F_{l-n}^{+}\,,\,W^{-}_{l-n} \]\,,\, F_{m}^{-} \],\\
\[E_{l}\,,\, b^{-1}F_{l-m}^{-} b \] &= &  \[E_{l}\,,\,F_{l-m}^{-}\] - \frac{1}{16} \[\sum_{n}  g_{mn} \[ F_{l-n}^{-}\,,\,W^{+}_{l-n} \]\,,\, F_{m}^{+} \],
\label{map3}
\\
\label{map4}
F^{\pm}_{m}&=& \mp  [E_{\pm l}\,,\, W_{l-m}^{\mp}],\\
\label{sumconst1}
\sum_{n=1}^{l-m-1} \[F_{n+m}^{\pm}\,,\, b^{\mp}F_{n}^{\mp}b^{\pm}\]&=&0,
\er
where the Eqs. (\ref{sumconst1}) are the constraints (\ref{sumconst}) written in a compact form. The relationships (\ref{map1})-(\ref{map4}) supplied with the conditions (\ref{isomor}) when conveniently substituted into Eqs. (\ref{eq1}) and (\ref{eq2})-(\ref{eq3}) decouple the L-ATM equations of motion into the GSG (\ref{gsg}) and GMT (\ref{eqt1})- (\ref{eqt2}) equations of motion, respectively.

The Noether and topological currents equivalence must also be considered along with the relationships (\ref{map1})- (\ref{sumconst1}), so the Eqs. (\ref{e1})-(\ref{e2}) written in the form (set $\zeta=1$)
\br
\label{equivalence11}
\epsilon^{\mu \nu}\mbox{Tr} \( E_{0} b^{-1}\pa_{\nu} b \) \,=\, 4\mbox{Tr}\(E_{0} \sum_{m} j^{\mu}_{m} \)
\er
where the currents components $j^{\mu}_{m}$ are given in Eqs. (\ref{u1})-(\ref{u11}), must be considered as an additional relationship between the fields of the GMT and GSG theories.  

Using (\ref{map2})-(\ref{map3}) and the constraints (\ref{sumconst1}) one can establish a linear system of equations for some spinor bilinears with ``coefficients'' given by certain exponentials of the Toda fields, and by solving this system of equations one can obtain certain spinor bilinears of the GMT model (\ref{gmt}) related to relevant $b$ fields of the GSG model (\ref{gsg1})-(\ref{gsg}), in a rather peculiar way. These relationships are analogous to those found in \cite{park} for fermion mass bilinears and bosonic ones for massive (free) non-Abelian fermions which are bosonized to certain symmetric space sine-Gordon models. 
The spinor bilinears expressed in terms of exponentials of Toda fields obtained in this way when conveniently substituted into  (\ref{map1}) will provide us some relationships between the parameters of both theories, i.e. the couplings $g_{mn}$, the fermion mass parameters of the GMT model (\ref{gmt}) and  the $\L_{m}^{\pm}$ parameters of the GSG theory (\ref{gsg1})-(\ref{gsg}). In fact, the various coupling parameters relationships will determine the appearance of certain kind of dualities between both theories. In the next sections we provide explicit examples for the assertions above.

\section{GSG/GMT relationship and the principal gradation of an affine Lie algebra}
\label{ssec:principal}

A model possessing the structures presented above is constructed in the following.
Let $\cgh$ be any (untwisted) affine Lie 
algebra provided with the principal gradation, where ${\bf s}=(1,1,\ldots
,1)$, and grading operator, $Q_{{\rm ppal}}$, given by
(\ref{grad}) with  $\ns = l \equiv  \mbox{\rm Coxeter number}$.
Therefore
\br
\cgh_0 &=& \{ H_a^0 \, , a=1,2, \ldots r\, ;\,\, C ;\,\, Q_{{\rm ppal}} \} \nonu\\
\cgh_p &=& \{ E_{\a(p)}^0,\,\,  E_{-\a(l-p)}^{1} \} \nonu\\
\cgh_{-p} &=& \{ E_{-\a(p)}^0, \,\, E_{\a(l-p)}^{-1} \}
\er
where $0<p<l$, and $\a(p)$  are positive roots of height $p$, i.e. they
 contain $p$  simple roots in their expansion.
In this gradation
 the dimension of $\cgh_0$ is minimal and this sub-algebra turns out to be Abelian with $\hat{{\cal G}}_{0}^{*}$ being the Cartan sub-algebra of ${\cal G}_{0}$.

Following (\ref{para}) we parameterize $b$ as
\be
\label{b}
b = e^{\vp \cdot H^0} 
\ee
We choose $E_{\pm l}$ to be
\be
\label{el}
E_{\pm l} = {\bf m} \cdot H^{\pm 1}
\lab{epmh}
\ee
where ${\bf m}$ is a vector inside the fundamental Weyl
chamber (FWC), and so ${\bf m}\cdot \a >0$ for $\a >0$. Then 
\be
E_0 \equiv {\bf m} \cdot H^{0}
\ee
satisfies (\ref{e0}) (with $\zeta =1$) since
$\lb E_0 \, , \, \cgh_0 \rb =0$.
Since the masses are determined by the eigenvalues of
$\lb E_{l}\, , \lb E_{-l} \, , \cdot \rb\rb$ \cite{matter} we conclude
that the $\vp$'s fields are massless, and the fields in the direction
of the step operators have masses $m_{\a}^2 =  4 ( {\bf m} \cdot \a )^2$.
In addition, since ${\bf m}$ lies inside the FWC, one has $m_{\a} \neq 0$
for any $\a$.
 
Therefore, according to the discussion in \cite{matter}, all the fields
in the direction of the step operators are Dirac fields. Then one can write
\br
\label{fspin1}
F^{+}_q = \sum_{\a(q)} \sqrt{im_{\psi^{\a(q)}}}
\psi^{\a(q)}_R \, E_{\a(q)}^0 +
\sum_{\a(l-q)} \sqrt{im_{\psi^{\a(l-q)}}}
{\tilde \psi}^{\a(l-q)}_R \, E_{-\a(l-q)}^1,
\er
\br
F^{-}_{l-q} = \sum_{\a(q)} \sqrt{im_{\psi^{\a(q)}}}
\psi^{\a(q)}_L \, E_{\a(q)}^{-1} -
\sum_{\a(l-q)} \sqrt{im_{\psi^{\a(l-q)}}}
{\tilde \psi}^{\a(l-q)}_L \, E_{-\a(l-q)}^0
\label{fspin2}
\er
with $0<q<l$, and
\be
m_{\psi^{\a(q)}} = m_{{\tilde \psi}^{\a(q)}} = 2 {\bf m} \cdot \a(q).
\ee

Thus, associated to the positive roots $\a(p)$ and  $\a(l-p')$, we have the Dirac fields
\br
\label{spinors}
\psi^{\a(p,i)} \equiv \(
\begin{array}{c}
\psi^{\a(p,i)}_R \\
\psi^{\a(p,i)}_L
\end{array} \)
\, ; \qquad
{\tilde \psi}^{\a(l-p', j)} \equiv \(
\begin{array}{c}
{\tilde \psi}^{\a(l-p', j)}_R \\
{\tilde \psi}^{\a(l-p', j)}_L
\end{array} \),
\er
where the additional index `\,$i$\,' denotes the $i$'th positive root $\a(p,i)$ with height $p$. According to the reality conditions (\ref{isomor}) the spinors (\ref{spinors}) are related to each other by complex conjugation
\br
\(\psi^{\a(p,i)}_{R,\, L} \)^{*}\,=\,{\tilde \psi}^{\a(l-p',j)}_{R,\,L},\,\,\,\,\,\,\,\,\,\,\,\,  l=p+p', \,\,\,\, \,i=j.
\er

In the following paragraphs we present some commutators which are useful to verify the relationships (\ref{map1})-(\ref{map3}). We have   
\br
\label{commutator}
[\vp.H^{0}\,,\,E_{\a}^{p}]\,=\, \vp_{a}\sum_{b=1}^{r} n^{\a}_{b} K_{ba} E_{\a}^{p},\,\,\,\,\mbox{where}\,\,\,\, \a = \sum_{b} n^{\a}_{b} \a_{b}.
\er

Using (\ref{commutator}) and the fact that $e^{T} F e^{-T}=F+[T,F]+... $, we have
\br
\nonu
[F_{q}^{-}\,,b F_{q}^{+} b^{-1} \,]&=&i \{ \sum_{\a(l-q,i)}  m_{\psi^{\a(l-q, i)}} \psi^{\a(l-q, i)}_L {\tilde \psi}^{\a(l-q, i)}_R \mbox{exp} \(- n_{b}^{\a(l-q, i)} K_{ba} \vp_{a}\) l_{c}^{\a(l-q, i)}\\
\label{sumfb}
&&+ \sum_{\a(q, i)}  m_{\psi^{\a(q, i)}} {\tilde \psi}^{\a(q, i)}_L \psi^{\a(q, i)}_R \mbox{exp} \( n_{b}^{\a(q, i)} K_{ba} \vp_{a}\) l_{c}^{\a(q, i)}\} H_{c}^{0}
\er
where $c$ is summed over $c=1,..,r$.

The commutator related to the GSG equation of motion becomes
\br
\nonu
[\L_{p}^{-}\,,b \L_{p}^{+} b^{-1} \,]&=&i \{ \sum_{\a(l-p, i)}  m_{\psi^{\a(l-p,i)}} \bar{\L}^{\a(l-p, i)} \mbox{exp} \(- n_{b}^{\a(l-p, i)} K_{ba} \vp_{a}\) l_{c}^{\a(l-p, i)}\\
\label{sumfb1}
&&+ \sum_{\a(p, i)}  m_{\psi^{\a(p, i)}}  \L^{\a(p, i)} \mbox{exp} \( n_{b}^{\a(p, i)} K_{ba} \vp_{a}\) l_{c}^{\a(p, i)}\} H_{c}^{0}
\er
where $\bar{\L}^{\a(l-p, i)}$ and $\L^{\a(p, i)}$ are some (complex) constant parameters which are the components of the $\L_{p}^{\pm}$'s when similar structure to the $F^{\pm}_{p}$'s given in (\ref{fspin1})-(\ref{fspin2}) are assumed for them, so
\br
\label{lambdas}
\bar{\L}^{\a(l-p, i)}\,=\,  \L^{\a(l-p, i)}_L {\tilde \L}^{\a(l-p, i)}_R,\,\,\,\, \L^{\a(p, i)}\,=\, {\tilde \L}^{\a(p, i)}_L \L^{\a(p, i)}_R.
\er

Taking into account Eqs. (\ref{b}), (\ref{el}), (\ref{fspin1}) and (\ref{fspin2}) the relationships (\ref{map2})-(\ref{map3}) give rise to the following expressions 
\br
e^{- <n^{\a(p,i)} . \vec{\vp}>} &=&1 + < n^{\a(p,i)}. {\bf m}>^{-1} (\psi^{\a(p,i)}_L)^{-1} \psi^{\a(p,i)}_R  \sum_{q,\, \a (q)} 4i g_{pq} {\tilde \psi}^{\a(q)}_L  \psi^{\a(q)}_L <n^{\a(p,i)}. l^{\a(q)}>\nonu\\
\label{exp1}\\
\nonu
e^{<n^{\a(l-p,i)} .\vec{\vp}>} &=& 1-  < n^{\a(l-p, i)}. {\bf m}>^{-1}  (\tilde{\psi}^{\a(l-p, i)}_L)^{-1} \tilde{\psi}^{\a(l-p, i)}_R \sum_{q,\, \a (q)} 4i g_{pq} {\tilde \psi}^{\a(q)}_L  \psi^{\a(q)}_L. \\
&&. <n^{\a(l-p, i)}. l^{\a(q)}> \label{exp2}\\
e^{<n^{\a(l-p, i)} . \vec{\vp}>} &=& 1 -  < n^{\a(l-p, i)}. {\bf m}>^{-1} (\psi^{\a(l-p, i)}_R)^{-1} \psi^{\a(l-p, i)}_L \sum_{q,\, \a(q)} 4i g_{pq} {\tilde \psi}^{\a(q)}_R  \psi^{\a(q)}_R. \nonu\\
&&. <n^{\a(l-p, i)}. l^{\a(q)}> \label{exp3}
\\
e^{-<n^{\a(p, i)} . \vec{\vp}>}  &=& 1 +  < n^{\a(p, i)}. {\bf m}>^{-1} (\tilde{\psi}^{\a(p, i)}_R)^{-1} \tilde{\psi}^{\a(p, i)}_L \sum_{q,\, \a(q)} 4i g_{pq} {\tilde \psi}^{\a(q)}_R  \psi^{\a(q)}_R <n^{\a(p, i)}. l^{\a(q)}>\nonu\\ 
\label{exp4}
\er 
where $ <x.y> \, \equiv \, K_{ab}  x_{a} y_{b} $.

The constraints (\ref{sumconst1}) provide the relationships (set $\eta_{0}=0$)
\br
\nonu
\sum_{q=1}^{l-p-1}\Big[ \sum_{ \a(p,q; i,j)}\{ \sqrt{m_{\psi^{\a(l-p-q, i)}}m_{\psi^{\a(l-q, j)}}}\, \epsilon_{-i,j} (1+ q_{1}) \tilde{\psi}_{R}^{\a(l-p-n, i)} \psi_{L}^{\a(l-q, j)} e^{<n^{\a(l-q, j)}. \vp>}\}-\\
\nonu
\sum_{ \b(p,q; i,j)} \sqrt{m_{\psi^{\a(p+q, i)}}m_{\psi^{\a(q, j)}}}\, \epsilon_{i,-j} (1+ q_{2})
\psi_{R}^{\a(p+q, i)} \tilde{\psi}_{L}^{\a(q, j)} e^{<n^{\a(q, j)}. \vp>} \Big]\\
=0,\,\,\,\,\,\,\,\,\,\,\,\,\,\,\,\,\,\,\, \label{sumconst11}
\er
\br
\sum_{q=1}^{l-p-1} \sum_{ \g(p,q;i,j)} \sqrt{m_{\psi^{\a(l-p-q, i)}}m_{\psi^{\a(q, j)}}}\, \epsilon_{i,j} (1+ q_{3}) \psi_{L}^{\a(l-p-q, i)} \psi_{R}^{\a(q, j)} e^{<n^{\a(q, j)}. \vp>}\,
=0\,,\,\,\,\,\ \label{sumconst12}
\er
where the sums are over all the roots of type $\a(p , i)$ such that $\a(p,q; i,j),\,\,\b(p,q; i,j)$,\,and \,$\g(p,q;i,j)$ are also roots defined by:   
$\a(p,q; i,j)\equiv \a(l-q, j)-\a(l-p-q, i)$,\,\,$\b(p,q; i,j) \equiv \a(p+q, i)-\a(q, j)$,\,\,$\g(p,q;i,j)\equiv \a(l-p-q, i)+\a(q, j)$;\,\,the $\epsilon_{ij} \equiv \pm 1$ originate from the Lie algebra commutators; $q_{1}$, $q_{2}$ and   $q_{3}$ are the highest positive integers such that $[\a(l-q, j)+q_{1} \a(l-p-q, i)]$,\,\, 
 [$-\a(q, j)-q_{2} \a(q+ p, i)$]\, and\,  [$\a(q, j)-q_{3} \a(l-q- p, i)$] are roots, respectively. Additional relationships are obtained by taking the complex conjugates of (\ref{sumconst11}) and (\ref{sumconst12}), respectively.

The currents equivalence relation Eq. (\ref{equivalence11}) becomes 
\br
\label{equivalence12}
\epsilon^{\mu \nu} \pa_{\nu} {\bf m}. \vp \,=\, \sum_{\a(q) > 0} \frac{2}{\a^2} m_{\psi^{\a(q)}} \bar{\psi}^{\a(q)} \g^{\mu} \psi^{\a(q)}. 
\er

On the other hand, conveniently substituting the Eqs. (\ref{exp1})-(\ref{exp4}) and (\ref{sumconst11})-(\ref{sumconst12}) into Eqs. (\ref{sumfb})-(\ref{sumfb1}), in order to satisfy (\ref{map1}), we have a set of relationships between the parameters $\L^{\pm}_{q}$ \( i.e. the $\L^{\a(q, i)}$,\, $\bar{\L}^{\a(q, i)}$ parameters in Eq. (\ref{lambdas})\),\, $m_{\psi}^{\a(q)}$ and $g_{pq}$. In fact, these relationships will encode the weak-strong duality exchange of the various coupling parameters: $g \rightarrow 1/g$. Below we present explicit constructions for the cases $\hat{sl}(2)$ and $\hat{sl}(3)$.

\section{Examples}

\subsection{The example of $\hat{sl}(2)$}

The simplest $\hat{sl}(2)$  L-ATM theory was extensively studied from different points of view in Refs. \cite{nucl, nucl1, tension, annals, witten}. Here we present some well known results in order to illustrate our constructions above. 
In this case the theory contains the usual sine-Gordon (SG) and the massive Thirring (MT) models
describing the soliton/particle correspondence of its spectrum \cite{nucl, nucl1, witten}. The one-(anti)soliton solution
satisfies the reality conditions (\ref{isomor}), the remarkable SG and MT classical correspondence in which, apart from the Noether and topological currents equivalence (\ref{equivalence12}), MT spinor bilinear is related to the exponential of the SG field \cite{nucl}.

In this case the Eqs. (\ref{b}), (\ref{el}), (\ref{fspin1}) and (\ref{fspin2}) become, respectively
\br
\label{sl21}
b&=&e^{\vp H^{0}} ,\,\,\,\,\,\,\,\, E_{\pm 2} \equiv \frac{1}{4}m_{\psi} \, H^{\pm 1}\\
\label{sl22}
F^{+}_1 &=& \sqrt{i m_{\psi}}\( \psi_R\, E_+^0 +
\widetilde \psi_R E_-^1\) \,\,\,\,\, ,\,\,\,\,\,
F^{-}_1 = \sqrt{i m_{\psi}}\( \psi_L\, E_+^{-1} -
\widetilde \psi_L\, E_-^0 \).
\er

Let us assume that the $\L^{\pm}_1$'s  have similar structures to the $F^{\pm}_1$, then
\br
\label{sl23}
\L^{+}_1 &=& \sqrt{i m_{\psi}}\( \L_R\, E_+^0 +
\widetilde \L_R E_-^1\) \,\,\,\,\,  \mbox{and}\,\,\,\,\,
\L^{-}_1 = \sqrt{i m_{\psi}}\( \L_L\, E_+^{-1} -
\widetilde \L_L\, E_-^0 \).
\er

Defining $ W^{\mp}_{1}$ through 
$F_{1}^{\pm}\,=\, \mp [E_{\pm 2}\,,\, W^{\mp}_{1}]$ one can write the action (\ref{latm}) as 
\br
\nonu
\frac{1}{k}{\cal I}_{l=2}=I_{WZW}(b) + \int_{M}\{ \frac{1}{2}<[E_{-2}\,,\,W_{1}^{+}] \pa_{+}W_{1}^{+} >-\frac{1}{2} <[E_{2}\,,\,W_{1}^{-}] \pa_{+}W_{1}^{-} >+\\
<F_{1}^{-} b F_{1}^{+} b^{-1} > \}\label{latm2}.
\er
Taking into account the parameterizations (\ref{sl21})-(\ref{sl22}) and the reality conditions (\ref{isomor}) $\(\widetilde{F_{1}^{\pm}}=F_{1}^{\pm} $,  with $\epsilon = 1$\), i.e.  $\tilde{\psi}_{R, L}\,=\, -(\psi_{R, L})^{*}$ and  $\vp$ pure imaginary (so, $\vp \rightarrow i \vp$ with the new $\vp$ being real), from (\ref{latm2}) one gets the $\hat{sl}(2)$ L-ATM Lagrangian \footnote{Consider
$\gamma_0 = -i \(
\begin{array}{rr} 0&-1\\ 1&0
\end{array}\)$ , 
$\gamma_1 = -i \(
\begin{array}{rr}  0&1\\  1&0
\end{array}\)$, 
$\gamma_5 = \gamma_0\gamma_1 = \(
\begin{array}{rr} 1&0\\ 0&-1
\end{array}\)$}
\br
\frac{1}{k}{\cal L}_{l=2} =\frac{1}{4} \pa_{\mu} \vp \, \pa^{\mu} \vp  
- i  {\bar{\psi}} \gamma^{\mu} \pa_{\mu} \psi
+ m_{\psi}\,  {\bar{\psi}} \,
e^{2i\vp\,\gamma_5}\, \psi,
\lab{atmsl2}
\er
The Eqs. (\ref{exp1})-(\ref{exp4}) provide the relationships \cite{nucl}
\br
\label{classicalboso}
\psi _{R}\tilde{\psi}_{L}=\frac{m_{\psi}}{4i g}(e^{-2
i\varphi }-1),\,\,\,\,&&\psi _{L}
\tilde{\psi}_{R}=-\frac{m_{\psi}}{4i g}(e^{2i\varphi }-1),
\er
and the currents equivalence Eq. (\ref{equivalence12}) becomes
\br
\label{equiv2}
\bar{\psi}\gamma^{\mu }\psi&=&\h \epsilon^{\mu\nu}\pa_{\nu}\vp.
\er

Notice that the relations (\ref{sumconst11})-(\ref{sumconst12}) in this case are trivialy satisfied. The relationships (\ref{classicalboso}) and (\ref{equiv2}) have been verified for $N=1$ and $N=1, 2$ solitons, respectively \cite{nucl, nucl1}.

In order to relate the parameters $\L,\, \bar{\L}$, $ m_{\psi}$ and $g$ we use the equation (\ref{map1}), so we need
\br
\label{map11}
[F^{-}_{1}\,,\,bF^{+}_{1} b^{-1}]&=& i m_{\psi} \( \psi_{L} \tilde{\psi}_{R}  e^{-2i\vp} + \tilde{\psi}_{L} \psi_{R}  e^{2i\vp} \) H^{0}\\
\label{map12}
[\L^{-}_{1}\,,\,b\L^{+}_{1} b^{-1}]&=& i m_{\psi} \( \L e^{-2i\vp} + \bar{\L}  e^{2i\vp} \) H^{0},
\er\
where, $\L = \L_{L}\tilde{\L}_{R},\,\,\bar{\L}=  \tilde{\L}_{L} \L_{R}$. Then substituting the spinor bilinears (\ref{classicalboso}) into (\ref{map11}) and comparing with (\ref{map12}) in accordance with (\ref{map1}), one gets 
\br
\label{strongweak22}
\L\, =\, \frac{m_{\psi}}{4i\, g}\,\, ,\,\,\,\,\,\,\,\,\bar{\L} = (\L)^{*}.
\er

The GSG Lagrangian (\ref{gsg1}) and Eq. (\ref{sl21}) provide $\frac{1}{k} {\cal L}_{SG}=\frac{1}{4}(\partial_{\mu}\vp)^{2} 
        +2  m_{\psi} |\L| \; \mbox{cos} 2 \vp$, which is just the usual sine-Gordon theory. On the other hand, the GMT Lagrangian (\ref{gmt}) for the fields (\ref{sl22}) becomes $\frac{1}{k}{\cal L}_{MT}= i\overline{\psi}\gamma^{\mu}\pa_{\mu}\psi-m_{\psi}\overline{\psi}\psi+ \frac{1}{2} g j_{\mu}j^{\mu}$, i.e. the usual massive Thirring model. Notice the weak-strong exchange $g \rightarrow \frac{1}{g}$ of the coupling constant.

\subsection{The example of $\hat{sl}(3)$}

The $\hat{sl}(3)$ ATM model has been considered in Refs. \cite{jmp, bueno}. In \cite{bueno} the soliton type solutions, calculation of their masses and
  the time delays due to the soliton interactions, as well as a confinement mechanism present in the model have been addressed. Whereas in \cite{jmp} the FJ method and an extension of it were applied to the  $\hat{sl}(3)$ L-ATM model in order to uncover its dual phases. In this subsection we re-derive the results of  \cite{jmp} in the full Lie algebraic constructions presented in previous sections.  
For the $\hat{sl}(3)$  L-ATM the Eqs. (\ref{sumconst11})-(\ref{sumconst12}) become non-trivial and the spinor bilinears have to be obtained in terms of the exponentials of the Toda fields by solving a linear system of equations as mentioned in the last paragraph of section 6.
 
The Eqs. (\ref{b}), (\ref{el}), (\ref{fspin1}) and (\ref{fspin2}) become, respectively
\br  
\label{e3}
b &=&e^{i\vp_{1} H^{0}_{1}+i\vp_{2} H^{0}_{2} },\\
\label{el1}
E^{\pm 3}&\equiv& {\bf m}.H^{\pm}\, =\,\frac{1}{6}[(2m^{1}_{\psi}+m^{2}_{\psi})H^{\pm 1}_{1}+(2m^{2}_{\psi}+m^{1}_{\psi})H^{\pm 1}_{2}]\\
\label{F1}
F_{1}^{+}&=&\sqrt{im^{1}_{\psi}}\psi _{R}^{1}E_{\alpha _{1}}^{0}+\sqrt{im^{2}_{\psi}}\psi
_{R}^{2}E_{\alpha _{2}}^{0}+\sqrt{im^{3}_{\psi}}\widetilde{\psi }_{R}^{3}E_{-\alpha
_{3}}^{1},
\\
\label{F2}
F_{2}^{+}&=&\sqrt{im^{3}_{\psi}}\psi _{R}^{3}E_{\alpha _{3}}^{0}+\sqrt{im^{1}_{\psi}}
\widetilde{\psi }_{R}^{1}E_{-\alpha _{1}}^{1}+\sqrt{im^{2}_{\psi}}\widetilde{\psi }
_{R}^{2}E_{-\alpha _{2}}^{1},
\\
\label{F3}
F_{1}^{-}&=&\sqrt{im^{3}_{\psi}}\psi _{L}^{3}E_{\alpha _{3}}^{-1}-\sqrt{im^{1}_{\psi}}
\widetilde{\psi }_{L}^{1}E_{-\alpha _{1}}^{0}-\sqrt{im^{2}_{\psi}}\widetilde{\psi }
_{L}^{2}E_{-\alpha _{2}}^{0},
\\
\label{F4}
F_{2}^{-}&=&\sqrt{im^{1}_{\psi}}\psi _{L}^{1}E_{\alpha _{1}}^{-1}+\sqrt{im^{2}_{\psi}}\psi
_{L}^{2}E_{\alpha _{2}}^{-1}-\sqrt{im^{3}_{\psi}}\widetilde{\psi }
_{L}^{3}E_{-\alpha _{3}}^{0}.
\er

Since the parameters $\L_{q}^{\pm}$ [as used in (\ref{sumfb1})-(\ref{lambdas})] have similar structures to the $F^{\pm}_{q}$'s we may write them simply by substituting $ \psi_{R,L}^{i} \rightarrow  \L_{R,L}^{i}$ and  $ \tilde{\psi}_{R,L}^{i} \rightarrow  \bar{\L}_{R,L}^{i}$ in Eqs (\ref{F1})-(\ref{F4}).

 Defining $ W^{\mp}_{q}$ through 
$F_{q}^{\pm}\,=\, \mp [E_{\pm 3}\,,\, W^{\mp}_{q}],\,\,q=1,2$, one can write the action (\ref{latm}) as 
\br
\nonu
\frac{1}{k}{\cal I}_{l=3}&=&I_{WZW}(b) + \int_{M} \sum_{q=1}^{2} \{ \frac{1}{2} <[E_{-3}\,,\,W_{q}^{+}] \pa_{+}W_{3-q}^{+} >-\frac{1}{2} <[E_{3}\,,\,W_{q}^{-}] \pa_{+}W_{3-q}^{-} >+\\
&&<F_{q}^{-}\,,\, b F_{q}^{+} b^{-1} > \}\label{latm3}.
\er

Using the parameterizations (\ref{e3})-(\ref{F4}) and the reality conditions (\ref{isomor})  $\( \widetilde{F_{1}^{\pm}}=F_{2}^{\pm}$,  with $\epsilon = 1$\),  i.e.\, $\tilde{\psi}^{i}_{R, L}\,=\, -(\psi^{i}_{R, L})^{*}$ and  $\vp_{a}$ pure imaginary (changing as usual $\vp_{a} \rightarrow i \vp_{a}$) from (\ref{latm3}) one gets the $\hat{sl}(3)$ L-ATM Lagrangian \cite{jmp}
\br
\label{atm3}
\frac{1}{k}{\cal L}_{l=3} = \sum_{\a(q)} \[ \frac{1}{24}\(\partial_{\mu }\phi_{\a(q)}\)^2 + i\overline{\psi}^{\a(q)} \gamma ^{\mu}\partial_{\mu }\psi^{\a(q)} - m_{\psi^{\a(q)}} \overline{\psi}^{\a(q)} e^{i \phi_{\a(q)}  \gamma_{5}}\psi^{\a(q)}\] 
\er
where $\phi_{1}=\a_{1}.\vp=2\vp_{1}-\vp_{2},\,\phi_{2}=\a_{2}.\vp=2\vp_{2}-\vp_{1},\,\phi_{3}=\a_{3}.\vp=\vp_{1}+\vp_{2}$,\,  $\a_{3}=\a_{1}+\a_{2}$.

The currents equivalence Eq. (\ref{equivalence12}) becomes
\br
\label{equivalence3}
\sum_{j=1}^{3} m^{j}_{\psi} \bar{\psi}^{j}\gamma^{\mu}\psi^{j} \equiv \epsilon^{\mu \nu}\partial_{\nu} (m^{1}_{\psi}\varphi_{1}+m^{2}_{\psi}\varphi_{2}),\,\,\,\,\,\,\, m^{3}_{\psi}=m^{1}_{\psi}+ m^{2}_{\psi},\,\,\,\,m^{i}_{\psi}>0.
\er

The equivalence (\ref{equivalence3}) has recently been verified for the various soliton species up to $2-$soliton \cite{bueno}. Moreover, the soliton solutions satisfy the reality conditions for the fields (\ref{isomor}), and the constraints (\ref{sumconst11})- (\ref{sumconst12}).

For $b$ and $E_{\pm 3}$ given in (\ref{e3})-(\ref{el1}) and taking into account the elements $\L^{\pm}_{1, 2}$ the Eq. (\ref{gsg1}) provides the GSG model 
\be
\label{sine3}
\frac{1}{k}{\cal L}_{GSG}[\vp]=\sum_{j=1}^{3}\[\frac{1}{24} \partial_{\mu}\phi_{j} \partial^{\mu}\phi_{j}
        +  \;2 m^{j}_{\psi} |\L^{j}| \mbox{cos} \phi_{j}\].
\ee

On the other hand, replacing (\ref{el1})-(\ref{F4}) into (\ref{gmt}) one obtains the GMT Lagrangian
\begin{equation}
\label{thirring3}
\frac{1}{k} {\cal L}_{GMT}[\psi,\overline{\psi}]= \sum_{j=1}^{3}\{i\overline{\psi}^{j}\gamma^{\mu}\pa_{\mu}\psi^{j} 
        - m^{j}_{\psi}\,\,{\overline{\psi}}^{j}\psi^{j}\}\,
        \,+ \frac{1}{2} \sum_{k,l=1}^{3} \[g_{kl} j_{k}.j_{l}\], 
\end{equation}
where $ j_{k}^{\mu}= \bar{\psi}^{k}\gamma^{\mu}\psi^{k}$.

From the relationships (\ref{map2})-(\ref{map3}) and (\ref{sumconst1}) between the GMT and GSG fields the bilinears of the type $\psi^{\a(m)}_{L}\tilde{\psi}_{R}^{\a(m)} $ and their complex conjugates deserve a special attention since they appear in the Eq (\ref{sumfb}) and is used to establish certain  relationships between the coupling constants of both models by comparing (\ref{sumfb}) and (\ref{sumfb1}) in accordance with Eq. (\ref{map1}). So, a system of linear equations for these kind of spinor bilinears can be written using the Eqs. (\ref{exp1})-(\ref{exp4}) and (\ref{sumconst11})-(\ref{sumconst12}). Solving such system of linear equations one may get 
\begin{eqnarray}
\nn
\frac{\psi _{R}^{1}\widetilde{\psi}_{L}^{1}}{i} &=&\frac{1}{4\D}[\left(
m^{1}_{\psi} p_{1}-m^{3}_{\psi}p_{4}-m^{2}_{\psi}p_{5}\right) e^{i\left( \varphi _{2}-2\varphi_{1}\right) }+m^{2}_{\psi}p_{5}e^{3i\left( \varphi _{2}-\varphi _{1}\right) }  \\
&&+m^{3}_{\psi}p_{4}e^{-3i\varphi _{1}}-m^{1}_{\psi}p_{1}] \label{duality31}\\
\nn
\frac{\psi _{R}^{2}\widetilde{\psi}_{L}^{2}}{i} &=&\frac{1}{4\D}[\left(
m^{2}_{\psi}p_{2}-m^{1}_{\psi}p_{5}-m^{3}_{\psi}p_{6}\right) e^{i\left( \varphi _{1}-2\varphi_{2}\right) }+m^{1}_{\psi}p_{5}e^{3i\left( \varphi _{1}-\varphi _{2}\right) } 
\\
&&m^{3}_{\psi}p_{6}e^{-3i\varphi _{2}}-m^{2}_{\psi}p_{2}]
\label{duality32} \\
\nn
\frac{\widetilde{\psi}_{R}^{3}\psi_{L}^{3}}{i} &=&\frac{1}{4\D}[\left(
m^{3}_{\psi}p_{3}-m^{1}_{\psi}p_{4}+m^{2}_{\psi}p_{6}\right) e^{i\left( \varphi_{1}+\varphi_{2}\right) }+m^{1}_{\psi}p_{4}e^{3i\varphi _{1}}  \\
&&-m^{2}_{\psi}p_{6}e^{3i\varphi _{2}}-m^{3}_{\psi}p_{3}],  \label{duality33}
\end{eqnarray}
where $\D \equiv g_{11}g_{22}g_{33}+2g_{12}g_{23}g_{13}-g_{11}\left(
g_{23}\right)^{2}-\left( g_{12}\right)^{2}g_{33}-\left( g_{13}\right)
^{2}g_{22}$;\,  $p_{1}\equiv \left( g_{23}\right)^{2}-g_{22}g_{33}$;\, $p_{2}\equiv \left( g_{13}\right) ^{2}-g_{11}g_{33}$;\, $p_{3}\equiv \left(
g_{12}\right)^{2}-g_{11}g_{22}$;\, $p_{4}\equiv g_{12}g_{23}-g_{22}g_{13}$;\, $p_{5}\equiv g_{13}g_{23}-g_{12}g_{33}$;\, $p_{6}\equiv g_{11}g_{23}-g_{12}g_{13}$.

Moreover, substituting the relations (\ref{duality31})- (\ref{duality33}) into  (\ref{sumfb}) one can show that the GSG parameters\, $\L^{j},\,\,\bar{\L}^{j}$ in (\ref{sine3}), the GMT couplings $g_{mn}$ and the mass parameters $m^{i}_{\psi}$ in (\ref{thirring3}), in order to satisfy (\ref{map1}), are related by
\begin{eqnarray}
\L^{1} &=&\frac{1}{4i\D} \[ m^{3}_{\psi}(g_{12}g_{23}-g_{13}g_{22})+ m^{1}_{\psi}(g_{22}g_{33}-g_{23}^2)\] , \label{strongweak31} \\
\L^{2} &=&\frac{1}{4i\D} \[ m^{3}_{\psi}(g_{12}g_{13}-g_{23}g_{11})+ m^{2}_{\psi}(g_{11}g_{33}-g_{13}^2)\] , \label{strongweak32} \\
\L^{3} &=&\frac{1}{4i\D} \[-\frac{m^{1}_{\psi}m^{2}_{\psi}}{(m^{3}_{\psi})}
(g_{12}g_{33}-g_{13}g_{23})+ m^{3}_{\psi} (-g_{11}g_{22}+(g_{12})^{2})\], \label{strongweak33}\\
\label{pps}
m^{3}_{\psi} p_{6} &=& -m^{1}_{\psi} p_{5},\,\,\,\,\,\,\,m^{3}_{\psi} p_{4}\, =\, -m^{2}_{\psi} p_{5},\,\,\,\,\,\,\,\,\, \bar{\L}^{i}\,=\,(\L^{i})^{*}.
\end{eqnarray}

Various limiting cases of the relationships (\ref{duality31})-(\ref{duality33}) and (\ref{strongweak31})-(\ref{strongweak33}) are possible \cite{jmp}. These relationships incorporate each $\hat{sl}(2)$ ATM sub-model (particle/soliton) 
weak/strong coupling correspondences; i.e., the MT/SG relationship (\ref{classicalboso})-(\ref{equiv2}) and (\ref{strongweak22}).

The procedure presented so far can directly be extended to the L-ATM model for the affine Lie algebra $\hat{sl}(n)$ furnished with the principal gradation. The theory will be described by the scalar fields $\vp_{a}\, (a=1,...n-1)$ and the Dirac spinors $\psi^{\a_{j}}$, $\widetilde{\psi}^{\a_{j}}$, ($j=1,...N$; $N\equiv \frac{n}{2}(n-1)$ = number of positive roots\, $\a_{j}$\, of the simple Lie algebra $sl(n)$) related to the GSG and GMT models, respectively. From the point of view of its solutions, the one-(anti)soliton solution associated to the field\, $\phi_{j}=\a_{j}.\vp$\, ($\vp=\sum_{a=1}^{n-1}\vp_{a} \a_{a}$,\, $\a_{a}$= simple roots of $sl(n)$) corresponds  to each pair of Dirac fields ($\psi^{\a_{j}}$, $\widetilde{\psi}^{\a_{j}}$)\cite{matter}.

\section{Conclusions and discussions}

We have defined a real and local Lagrangian, the so-called L-ATM theory (\ref{latm}) imposing some reality conditions (\ref{isomor}) and constraints (\ref{sumconst}) on the fields of the off-critical ATM model defined by action (\ref{effaction1}) with $\eta=$const. 
We have shown, in the context of the master Lagrangian and Faddeev-Jackiw symplectic methods, that the L-ATM (\ref{latm}) theory associated to any (untwisted) affine Lie algebra under certain special conditions expressed in Eqs. (\ref{e0}) and (\ref{abelian}), is a master Lagrangian \cite{deser} from which both the GMT (\ref{gmt}) and the GSG (\ref{gsg1})-(\ref{gsg}) models are derivable. We have provided some mapping between the fields of the dual models (\ref{map1})- (\ref{equivalence11}) which decouples the L-ATM equations of motion (\ref{eq1})-(\ref{eq3}) into the equations of motion of each dual model, the Eqs. (\ref{gsg}) and (\ref{eqt1})-(\ref{eqt2}), respectively. We have presented an explicit construction of these relationships in the case of an affine Lie algebra furnished with the principal gradation. In the case of $\hat{sl}(n)\, (n=2, 3)$ we have provided the explicit constructions of the fields relationships and various duality exchanges of the coupling constants of type: $g \rightarrow 1/g$. 

The master Lagrangian approach used to recover the GSG model is more suitable than the symplectic methods since the problem of solving the spinors in terms of the scalars in the constraints (\ref{const2}) is intractable. In the previous works \cite{jmp, annals} on $\hat{sl}(n)\, (n=2,3)$ L-ATM reductions based on extensions of the FJ method, the so-called symplectic quantization methods, it was observed the loss of covariance in the GSG sector and even the group symmetry in the bosonic sector was hidden.

The L-ATM model allowed us to establish these type of relationships for interacting massive spinors in the spirit of particle/soliton correspondence. We hope that the quantization of the L-ATM theories and the related WZNW models, and  in particular the relationships of type (\ref{exp1})-(\ref{exp4}), would provide the bosonization dictionary of the GMT fermions in terms of their associated Toda and WZNW fields. In addition, the above approach to the GMT/GSG duality may be useful to construct the conserved currents and the algebra of their associated charges in the context of the CATM $\rightarrow$ L-ATM reduction. These currents in the MT/SG case were constructed treating each model as a perturbation on a conformal field theory (see \cite{kaul} and references therein).

Notice that in the GSG model (\ref{gsg1})-(\ref{gsg}) the WZ-term does not contribute for ${\cal G}_{0}^{*}$ Abelian according to (\ref{abelian}); i.e. $b$ is an element of an Abelian group. In relation to this point, let us mention that in \cite{rajeev} it has been conjectured that the nonlinear sigma model without the WZ term for a compact simple group is equivalent as a quantum theory to the so-called non-Abelian massless Thirring model.

In particular, the $\hat{sl}(n)$ L-ATM models may be relevant in the construction of the low-energy effective theories of multi-flavor QCD$_{2}$ (in which the baryons of QCD$_{2}$ are sine-Gordon type solitons \cite{frishman}) with the dynamical fermions in the fundamental and adjoint representations, so providing an extension of the picture described in \cite{tension} for the $\hat{sl}(2)$ case.  In multi-flavor QCD$_{2}$ effective Lagrangian the ``baryons'' of the theory may be described by GMT-like theories as conjectured in \cite{gonzales}.

Moreover, two dimensional models with four-fermion interactions have played an important role in the understanding of QCD (see, e.g. \cite{bennett} and references therein). A $\hat{sl}(3)$ GMT sub-model with two-fermion species defines the so-called extended Bukhvostov-Lipatov model (BL) and has recently been studied by means of a bosonization technique \cite{sakamoto1}. BL type models were applied to $N-$body problems in nuclear physics \cite{sakamoto}. Fermion-boson duality in 1-space quantum mechanical many-body system with non-vanishing  zero-range interactions has been derived in \cite{cheon}, we believe that the GSG/GMT relationship could provide a field theoretical framework to study such type of problems.

Although S-duality must be seen at the quantum level (the 2D analogue of the Montonen-Olive conjecture) we hope that the present work provides the algebraic recipe to construct the bosonic dual of the GMT theory (\ref{gmt}) associated to any (untwisted) affine Lie algebra since the form of the relationships of type ({\ref{strongweak22}) and ({\ref{strongweak31})-({\ref{strongweak33}) between the couplings are expected to remain almost the same in the full quantum regime (in the simplest $\hat{sl}(2)$ case the SG/MT S-duality implies the shift of $g$ by a constant in Eq. (\ref{strongweak22}) \cite{coleman}).   

Notice that relaxing one or both of the conditions (\ref{abelian}) one can get some deformations of the GMT(or the GSG) model. For example the case $\hat{sl}(2)\, (l=3)$ contains the SG theory as a sub-model \cite{matter}. For some particular Lie algebra data the authors of Refs. \cite{sotkov1, fateev} have been concerned with such type of systems. In \cite{fateev} the author studied families of S-dual pairs of integrable models in which the strong phase is represented by (complex) sine-Gordon model interacting with $\hat{{\cal G}}$ Abelian affine Toda theories and the weak coupling models represent MT fermions interacting with the Abelian affine Toda model based on the dual algebra $\hat{{\cal G}}^{v}$.
 
Finally, using the explicit expression for the effective action (\ref{effaction1}) corresponding to the CATM model (\ref{eqn11})-(\ref{eqn41}) one can obtain the energy-momentum tensor, and hence compute in a direct way the masses of the solitons which were calculated in \cite{matter} as the result of the spontaneous breakdown of the conformal symmetry. Moreover, these masses should reproduce the ones obtained by simply using the corresponding terms of the dual model Lagrangians (\ref{gsg1}) and (\ref{gmt}), respectively.

{\sl Acknowledgements}

The author thanks Prof. A. Accioly for encouragement and the referee for valuable suggestions. This work was supported by a FAPESP grant.

\appendix

\section{Affine Lie algebras}
\label{app:a}
We record some Lie algebraic properties used in the present paper following the notations and conventions of Ref. \cite{laf}. Consider an (untwisted) affine Kac-Moody algebra $\hat{{\cal G}}$ furnished with an integral gradation  $\hat{{\cal G}}$  = $\bigoplus_{n \in \IZ} \hat{{\cal G}}_{n}$,\, and denote
\begin{eqnarray}
\label{pm}
\hat{{\cal G}}_{+}\, =\, \bigoplus_{n>0} \hat{{\cal G}}_{n};\,\,\,\,\hat{{\cal G}}_{-}\, =\, \bigoplus_{n>0} \hat{{\cal G}}_{-n}.
\end{eqnarray}

The commutation relations for an affine Lie algebra in the Chevalley basis are
\br
&&\left[ \emph{H}_a^m,\emph{H}_b^n\right] =mC\frac{2}{\alpha_{a}^2}K_{a b}\delta _{m+n,0}  \lab{a7}\\
&&\left[ \emph{H}_a^m,E_{\pm \alpha}^n\right] = \pm K_{\alpha a}E_{\pm \alpha}^{m+n} 
\lab{a8}\\
&&\left[ E_\alpha ^m,E_{-\alpha }^n\right]
=\sum_{a=1}^rl_a^\alpha \emph{H}_a^{m+n}+\frac 2{\alpha ^2}mC\delta
_{m+n,0}  \lab{a9}
\\
&&\left[ E_\alpha ^m,E_\beta ^n\right] =
\varepsilon (\alpha ,\beta ) (1+ q)E_{\alpha +\beta }^{m+n};\qquad \mbox{if }\alpha
+\beta \mbox{ is a root \qquad }  \lab{a10}
\\
&&\left[ D,E_\alpha ^n\right] =nE_\alpha ^n,\qquad \left[ D,\emph{H}%
_a^n\right] =n\emph{H}_a^n.  \lab{a12}
\er
where $K_{\alpha a}=2\a.\a_{a}/\a_{a}^2=n_{b}^{\a}K_{ba}$, with $n_{a}^{\a}$ and $l_a^\alpha$ being the integers in the expansions $\a=\sum_{a} n_{a}^{\a}\a_{a}$ and $\a/\a^2=\sum_{a} l_a^\alpha\a_{a}/\a_{a}^2$, and $\varepsilon (\alpha ,\beta )$ the relevant structure constants. $q$ is the highest positive integer such that $\b - q \a $ is a root.   

The symmetric non-degenerate bilinear form of $\hat{{\cal G}}$ is normalized as
\begin{eqnarray}
\label{tr}
Tr\( H_{a}^{m} H_{b}^{n} \), =\, \delta_{m+n, 0} \eta_{ab},\,\,\,\,\, Tr\(E_{\a}^{m} E_{\b}^{n} \), =\, \frac{2}{\a^2}\delta_{m+n, 0}\delta_{\a+\b, 0} \\
Tr\( C T_{a}^{m} \), =\,Tr\( D T_{a}^{m} \)\, =\,0; \,\,Tr\( C D \), =\,1.
\end{eqnarray}

The integral gradations of ${\cal G}$ are defined by a set of co-prime integers ${\bf s}= \(s_{0},\,s_{1},... s_{r}\)$ and grading operators given by \cite{kac}
\begin{eqnarray}
\label{grad}
Q_{{\bf s}}\equiv H_{{\bf s}}+N_{{\bf s}}D-\frac{1}{2N_{{\bf s}}}Tr(H_{{\bf s}
}^{2}) C
\end{eqnarray}
where
\begin{eqnarray}
\label{graddef}
H_{{\bf s}} = \sum_{a=1}^{r} s_{a} \lambda_{a}^{r}. H^{0};\,\,\,\, N_{{\bf s}} = \sum_{i=0}^{r} s_{i} m_{i}^{\psi}.
\end{eqnarray}
The $H^{0}$'s are the Cartan sub-algebra generators of ${\cal G}$; $\alpha_{a}, a=1,2,...,r$ are its simple roots; $\psi$ is its maximal root; $m_{a}^{\psi}$
are the integers in the expansion $\psi= \sum_{a=1}^{r} m_{a}^{\psi} \alpha_{a},\,\, m_{0}^{\psi} \equiv 1$; and $\lambda_{a}^{v}$ are the fundamental co-weights satisfying the relation $\alpha_{a}.\lambda_{b}^{v}= \delta_{ab}$.

Notice from (\ref{grad})-(\ref{graddef}) that the generators of $\hat{{\cal G}}_{0}$ are\,
 $\{ C,D, H_{a\,(a=1,2,...,r)}^{0}, E_{\pm \alpha}^{0},\,E_{\pm \b}^{\mp 1}\}$,  such that $\sum_{a=1}^{r}s_{a} \lambda_{a}^{v}. \a =0 $
 and $\sum_{a=1}^{r}s_{a} \lambda_{a}^{v}. \b =N_{{\bf s}}$. Let us denote by $\hat{{\cal G}}_{0}^{*}$ the sub-algebra of  $\hat{{\cal G}}_{0}$ generated by all elements of $\hat{{\cal G}}_{0}$ except $ Q_{{\bf s}}$ and $C$, then we have $[\hat{{\cal G}}_{0}^{*}\,,\,Q_{{\bf s}}]=[\hat{{\cal G}}_{0}^{*}\,,\,C]=0$.

Define a basis in which all generators of $\hat{{\cal G}}_{0}$ are orthogonal to  $ Q_{{\bf s}}$ and $C$ \cite{laf}
\br
\label{defining}
\widetilde{H}^{0}_{a}\, =\, H^{0}_{a}-\frac{1}{N_{{\bf s}}} Tr\(H_{{\bf s}} H_{a}^{0}\)C = H^{0}_{a} -\frac{2 s_{a}}{ \a_{a}^2\, N_{{\bf s}}} C
\er
then we have
\begin{eqnarray}
Tr(C^{2}) &=&Tr(C\widetilde{H}^{0}_{a})=Tr(Q_{{\bf s}}^{2})=Tr(Q_{{\bf s}}
\widetilde{H}^{0}_{a})=0,\,\,Tr(Q_{{\bf s}}C) = N_{{\bf s}}  \nonumber \\
Tr(\widetilde{H}^{0}_{a}\widetilde{H}^{0}_{b})&=&Tr(H^{0}_{a}H^{0}_{b})= 4 \frac{\a_{a}.\a_{b}}{\a_{a}^2  \a_{b}^2} \equiv \eta_{ab}, \,\,\,\, a,b= 1,2,...,r
\end{eqnarray}
where ${H}^{0}_{a}= 2\frac{\a_{a}.H^{0}}{\a_{a}^2}$,\,\, $Tr (x.H^{0} y.H^{0})= x.y$.
 
For $B$ given in (\ref{para}) the following relations hold 
\br
\label{identity1}
\epsilon^{ijk} Tr\( B^{-1} \pa_{i}B  B^{-1} \pa_{j}B  B^{-1} \pa_{k}B\)& =& \epsilon^{ijk} Tr\( b^{-1} \pa_{i}b  b^{-1} \pa_{j}b  b^{-1} \pa_{k}b\)\\
\label{identity2}
Tr\( \pa_{\mu}B \pa^{\mu} B^{-1}\) &=& Tr ( \pa_{\mu}b \pa^{\mu} b^{-1}) - 2N_{{\bf s}}  \pa_{\mu} \nu \pa^{\mu} \eta,
\\
\label{identity3}
\,\,\,\,Tr\(F^{+}_{m} B^{-1}F^{-}_{m} B \)&=& e^{m \eta} Tr \(F^{+}_{m} b^{-1}F^{-}_{m} b \)
\er

\end{document}